%% file: paper.tex
\documentclass[twocolumn]{aastex701}

\usepackage{amsmath,amssymb,amsfonts}
\usepackage{graphicx}
\usepackage{booktabs}
\usepackage{multirow}
\usepackage{bm}
\usepackage{threeparttable}

\shorttitle{The HINSPHERES Framework}
\shortauthors{Xing et al.}

\begin{document}

%%======================================================================
%% Title
%%======================================================================
\title{HINSPHERES: Spherically Symmetric Radiative Transfer Modeling of HINSA and the Radial HI-to-H$_2$ Transition in Molecular Clouds}

\author[0000-0002-3411-9654]{Yuchen Xing}
\affil{National Astronomical Observatories, Chinese Academy of Sciences, Beijing 100101, China}
\affil{University of Chinese Academy of Sciences, Beijing 100049, China}
\email{xingyc@bao.ac.cn}

\author[0000-0003-3010-7661]{Di Li}
\affil{National Astronomical Observatories, Chinese Academy of Sciences, Beijing 100101, China}
\affil{Department of Astronomy, Tsinghua University, Beijing 100084, China}
\email[show]{dili@tsinghua.edu.cn}

\author[0000-0001-8550-0095]{Xin Lyu}
\affil{National Astronomical Observatories, Chinese Academy of Sciences, Beijing 100101, China}
\affil{University of Chinese Academy of Sciences, Beijing 100049, China}
\email{lvxin@bao.ac.cn}

\author{Pei Zuo}
\affil{National Astronomical Observatories, Chinese Academy of Sciences, Beijing 100101, China}
\email{peizuo@nao.cas.cn}

\author{Yulian Guo}
\affil{National Astronomical Observatories, Chinese Academy of Sciences, Beijing 100101, China}
\affil{University of Chinese Academy of Sciences, Beijing 100049, China}
\email{guoyl@bao.ac.cn}

\author{Silu Xu}
\affil{National Astronomical Observatories, Chinese Academy of Sciences, Beijing 100101, China}
\affil{University of Chinese Academy of Sciences, Beijing 100049, China}
\email{xusl@bao.ac.cn}

\author{Zhijun Zhang}
\affil{National Astronomical Observatories, Chinese Academy of Sciences, Beijing 100101, China}
\affil{University of Chinese Academy of Sciences, Beijing 100049, China}
\email{zhangzj@bao.ac.cn}

\author{Jinru Li}
\affil{National Astronomical Observatories, Chinese Academy of Sciences, Beijing 100101, China}
\affil{University of Chinese Academy of Sciences, Beijing 100049, China}
\email{lijr@bao.ac.cn}

\author[0000-0003-3758-7426]{Yan Duan}
\affiliation{Space Engineering University, Beijing 101416, China}
\email{yduan@nao.cas.cn}

\author{Rouyu Li}
\affil{National Astronomical Observatories, Chinese Academy of Sciences, Beijing 100101, China}
\affil{University of Chinese Academy of Sciences, Beijing 100049, China}
\email{liry@bao.ac.cn}

\author[0009-0007-3201-9100]{Jiawei Liu}
\affil{National Astronomical Observatories, Chinese Academy of Sciences, Beijing 100101, China}
\affil{University of Chinese Academy of Sciences, Beijing 100049, China}
\email{liujw@bao.ac.cn}

\author{Jifeng Xia}
\affil{National Astronomical Observatories, Chinese Academy of Sciences, Beijing 100101, China}
\affil{University of Chinese Academy of Sciences, Beijing 100049, China}
\email{xjf@nao.cas.cn}

\author{Jinjin Xie}
\affil{State Key Laboratory of Radio Astronomy and Technology, Xinjiang Astronomical Observatory, Chinese Academy of Sciences, 150 Science 1-Street, Urumqi, Xinjiang 830011, China}
\email{xiejinjin@xao.ac.cn}

\begin{abstract}
The transition from atomic hydrogen (HI) to molecular hydrogen (H$_2$) is the first step in the evolution from diffuse interstellar gas to dense star-forming cores. HI narrow self-absorption (HINSA), produced by cold atomic gas mixed with molecular material, provides a direct tracer of the HI-to-H$_2$ transition in molecular clouds. Existing HINSA analyses typically model HINSA as a single Gaussian component with a uniform spin temperature. This approach cannot recover the radial HI distribution and yields large errors for young cores whose profiles deviate from a Gaussian shape. Here we present \texttt{hinspheres}, an open-source framework that models the core as onion-like spherically symmetric concentric shells, solves radiative transfer along the line of sight and jointly fits the HI abundance, temperature, and kinematic parameters. The framework recovers all parameters with a mean relative error of 4.6--6.8\% on our synthetic data with different ages. A combined resolution--noise grid shows that the median recovery error remains below 16\% for cores within 1\,kpc observed at FAST resolution with noise levels below $0.2$~K. Applying it to the dark cores L1574 and L1544, we reveal radial HI abundance profiles that span about three orders of magnitude, which were missed by the single-Gaussian method. The derived chemical ages are $6.5\times10^6$ and $8.7\times10^6$ yr, respectively. Compared with the commonly used single-Gaussian method, \texttt{hinspheres} characterizes the radial profile of HINSA structures more accurately, making it well suited to identifying and analyzing young cores undergoing the HI-to-H$_2$ transition.
\end{abstract}


%%======================================================================
%% 1. Introduction
%%======================================================================
\section{Introduction}
\label{sec:intro}

The conversion of atomic hydrogen (HI) to molecular hydrogen (H$_2$) is
the first step of the evolution from diffuse interstellar gas to dense
star-forming cores~\citep{Bergin2007, Tielens2005}.  In the standard
picture, a diffuse atomic cloud contracts and shields itself from the
interstellar radiation field.  Once self-shielding is sufficient, H$_2$
forms on grain surfaces and the cloud transitions from atomic to
molecular~\citep{Hollenbach1971, Draine2011}.  The timescale and spatial
structure of this HI-to-H$_2$ transition remain poorly constrained
observationally~\citep{Bergin2007, Goldsmith2007}, yet they set the
initial conditions for all subsequent star formation.

Tracing this conversion is difficult: H$_2$ lacks a permanent electric dipole moment, and its first rotational level lies far above the typical kinetic energy of particles in the cold gas. Carbon monoxide, the standard
surrogate tracer, has its own limitations: in the dense, well-shielded
interior CO freezes onto dust grains~\citep{Caselli1999}, while in the
translucent outer layers CO photodissociation creates large volumes of
``CO-dark'' molecular gas~\citep{Bolatto2013,Xia2026}. HI narrow self-absorption
(HINSA) usually shows up as a narrow absorption dip on the broader
21\,cm HI emission line. It is produced by cold ($T \lesssim 20$\,K) atomic
gas mixed with molecular material absorbing the warmer background HI
emission~\citep{Li2003}.  Note that whether this cold gas layer shows
absorption, emission, or no detectable feature at all is not an intrinsic
property of the gas itself but a radiative-transfer effect set by the
contrast between its spin temperature and the background brightness,
together with the line opacity~\citep{Li2003, Tielens2005}.  Because the
cold HI producing HINSA is physically co-spatial with the molecular gas
in which it is embedded, HINSA is a direct tracer of the HI-to-H$_2$
transition.

HINSA is commonly detected in cold dark clouds and is frequently
associated with molecular tracers such as CO, OH, and HC$_3$N.
\citet{Li2003} detected HINSA in 77\% of surveyed dark cores in the
Taurus/Perseus region and showed that HINSA correlates with the
accompanying OH and $^{13}$CO emission in position, velocity, and
non-thermal line width.  \citet{Krco2010} detected HINSA in 40 dark
clouds within 700\,pc and measured a mean projected HI abundance $f_{\rm HI,2D}$ of
$10^{-2.8}$, confirming that cold HI is typically the third most
abundant component in molecular clouds.  A pilot HI survey of 17
Planck Galactic cold clumps using FAST detected HINSA in 58\% of the
clumps~\citep{Tang2020}, and a subsequent FAST survey guided by
HC$_3$N emission achieved HINSA detection rates of 92\% overall and
87\% among clumps with HC$_3$N detections~\citep{Liu2022}.  HINSA has
also been found toward the dense ``bones'' filaments of the inner
Galaxy~\citep{Sun2024} as well as in CO-dark molecular gas in the
outer Galactic disk~\citep{Luo2024}.

The radial distribution of the HI abundance $f_{\rm HI}(r)$ is central to understanding the
chemical evolution of molecular clouds.  Time-dependent models of H$_2$
formation on grains show that the dense inner regions approach reaction
equilibrium faster than the diffuse outer envelope, so that at late
times the HI abundance rises toward the cloud edge, from
$\sim 10^{-4}$ in well-shielded cores to a much higher value at the
warm boundary~\citep{Goldsmith2007}.  Direct evidence for such radial structure was obtained by~\citet{Zuo2018}, who combined HINSA with CO, dust, and extinction measurements toward the isolated core B227. They revealed a ring of enhanced HI abundance, measured the projected HI abundance varying from 2\% to 0.2\% within the core, and derived a cloud formation timescale of $\sim 6\times10^6$~yr. The predicted radial
variation implies that, during the HI-to-H$_2$ transition, the
high-density core converts faster than the envelope and reaches a lower
HI abundance at the center. At equilibrium, the equilibrium HI abundance decreases with increasing density \citep{Goldsmith2005}, so the densest central region also attains the lowest HI abundance.  %Real HINSA profiles should therefore deviate from a simple Gaussian, \textbf{as the HI in the core center and the envelope yield different depths of self-absorption, which jointly modify the radiative transfer and shape the line profile along the line of sight.}TODO

HINSA is typically treated as a single absorbing component with uniform
spin temperature.  In the three-component framework
of~\citet{Li2003}, the observed HI brightness temperature is understood
as the line-of-sight result of background, cold, and foreground HI
layers, with the cold cloud in the middle producing the HINSA feature.
The foreground correction is estimated from a simplified model of the
Galactic HI distribution and the cloud distance, while the background
emission is reconstructed by fitting polynomials around the absorption
feature to isolate the HINSA component.  The HINSA profile is then
fitted with a single Gaussian to obtain the optical depth and column
density.  \citet{Krco2008} adopt a forward-fitting approach
with a second-order derivative optimization, modeling the line-of-sight
as a background component and a single cold-cloud component with a
Gaussian HINSA profile.  The optical depth is obtained through iterative
fitting.  In both approaches, a single
absorption profile cannot by itself separate the opacity from the
temperature, so the spin temperature must be assumed (typically
10\,K).  Subsequent HINSA studies have largely followed one of these
two frameworks.  Studies based on~\citet{Li2003} have experimented with
higher-order polynomial backgrounds and multi-Gaussian
decompositions~\citep{Tang2020, Liu2019, Li2022}, while
studies based on~\citet{Krco2008} have refined the forward-fitting
approach~\citep{Krco2010, Liu2022, Luo2024}.  Both approaches assume that the
HINSA profile is single-component Gaussian and cannot invert observed
spectra to recover the underlying radial HI distribution.  This is
especially problematic for cores undergoing the HI-to-H$_2$ transition,
whose HINSA profiles are expected to deviate from Gaussian and which are
key objects for studying the atomic-to-molecular conversion.

In this paper, we present \texttt{hinspheres}, a Python framework that
fills this gap.  The framework models spherically symmetric clouds with
parametric density and temperature profiles and a discrete radial HI
abundance distribution, performing layer-by-layer radiative transfer
along every line of sight through a three-dimensional data cube.
By fitting all lines of sight simultaneously, \texttt{hinspheres} constrains the full radial HI abundance distribution, as well as the spin temperature and opacity. Compared with the single-Gaussian method, which returns only one column-density value per line of sight and no radial information, \texttt{hinspheres} characterizes the radial profile of HINSA structures more accurately, making it well suited to identifying and analyzing young cores undergoing the HI-to-H$_2$ transition.

The paper is organized as follows.  Section~\ref{sec:model} describes
the physical model and radiative transfer calculation.
Section~\ref{sec:fitting} presents the data preparation and fitting
algorithm, and discusses the fundamental parameter degeneracies.  Section~\ref{sec:validation} validates the pipeline on synthetic data.
Section~\ref{sec:application} demonstrates an application to
observational data.  Section~\ref{sec:discussion} examines the non-Gaussianity of
HINSA absorption profiles, compares the \texttt{hinspheres} three-dimensional
fitting approach with the conventional single-Gaussian method, and
discusses the applicability and limitations of the framework.
Section~\ref{sec:summary} summarizes the results.

%%======================================================================
%% 2. Physical Model
%%======================================================================

%%======================================================================
%% 2. Physical Model + Radiative Transfer
%%======================================================================
\section{Physical Model and Radiative Transfer}
\label{sec:model}

\begin{figure*}[htbp]
  \centering
  \includegraphics[width=0.6\textwidth]{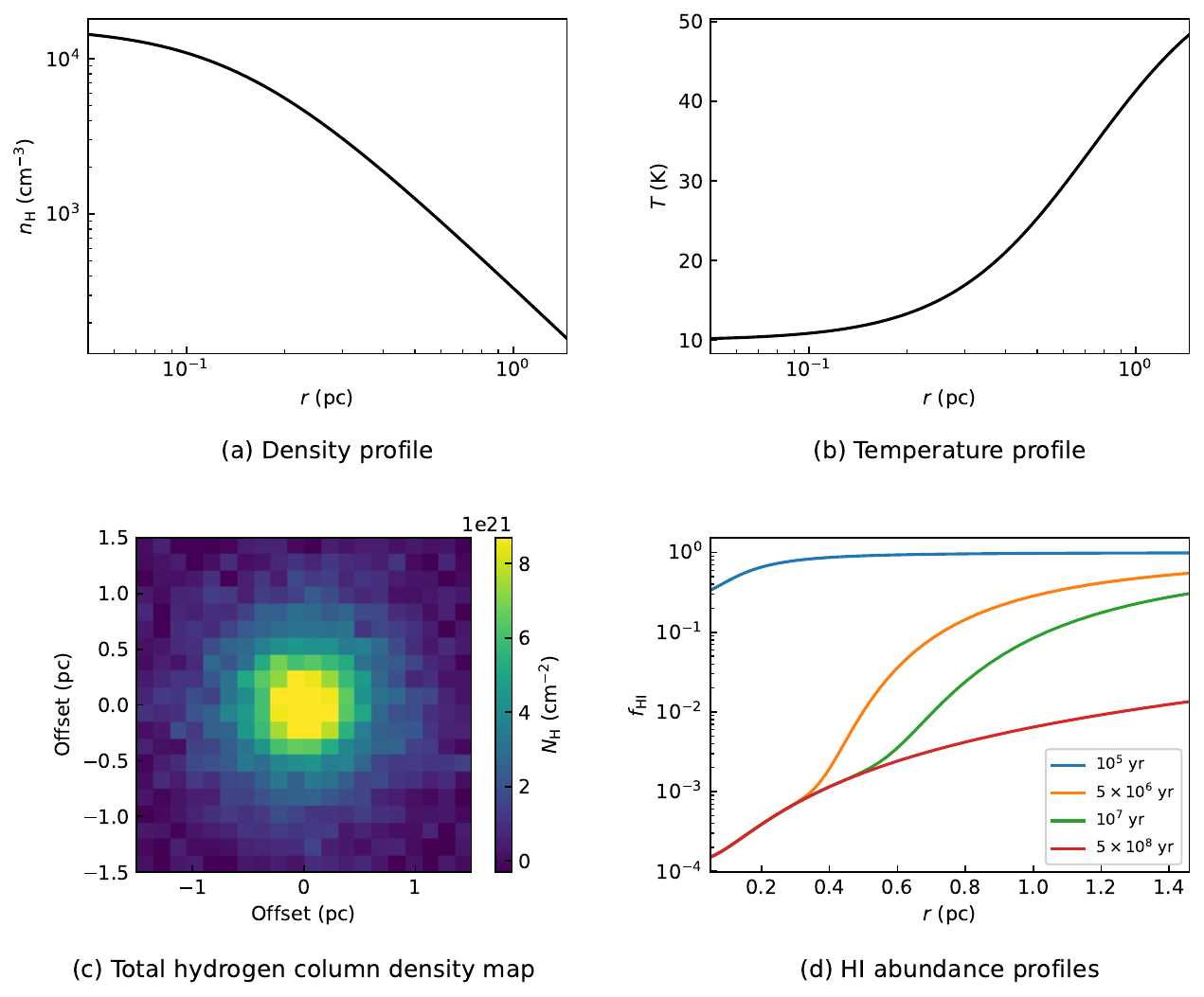}
  \caption{Profiles and $N_{\rm H}$ map of our synthetic core in
  Section~\ref{sec:synthetic}.  (a) Density profile for the synthetic core.
  (b) Temperature profile.  (c) $N_{\rm H}$ map generated from the
  density profile, convolved to a Planck-like resolution of
  $5\arcmin$ with a pixel scale of $1.72\arcmin$. Gaussian noise of
  $\sim 4\times 10^{20}$~cm$^{-2}$~pixel$^{-1}$ is added to mimic the
  Planck $\tau_{353}$ uncertainty.
  (d) HI abundance $f_\mathrm{HI}$ profiles at four chemical ages
  calculated from the time-dependent model of \citet{Goldsmith2007}.}
  \label{fig:radial_profiles}
\end{figure*}

The \texttt{hinspheres} model represents a spherically symmetric cloud
as $N$ concentric shells spanning the outer radius $R_\mathrm{out}$
(Figure~\ref{fig:radial_profiles}).  Shell $k$ ($k = 1, \ldots, N$)
occupies the radial interval $r_{k-1} \le r < r_k$ with $r_0 = 0$ and
outer boundaries $r_k = k\,R_\mathrm{out} / N$.  Physical quantities are
evaluated at the shell midpoints and taken as constant across each
shell.

For a line of sight at impact parameter $b$ from the cloud center, the
path length through shell $k$ is
\begin{equation}
  \delta l_k(b) = \sqrt{r_k^2 - b^2} - \sqrt{r_{k-1}^2 - b^2},
  \label{eq:path_length}
\end{equation}
where the second square-root term is set to zero if $b \geq r_{k-1}$.
Each intersected shell is traversed twice (back and front hemispheres),
including at the cloud center ($b = 0$), where the line of sight pierces
the full cloud diameter. 

We adopt a generalized Plummer profile for the density structure,
a form that has been widely used in core modeling
\citep{Tafalla2004, Marsh2014, Xing2026} because it captures
the flat central region and power-law outer envelope characteristic
of both Bonnor--Ebert spheres ($\alpha \approx 2.4$) and collapsing
power-law profiles (smaller $r_0$):
\begin{equation}
  n_{\rm H}(r) = \frac{n_0}{1 + (r / r_0)^\alpha},
  \label{eq:plummer_density}
\end{equation}
where $n_0$ is the central hydrogen-nucleus number density, $r_0$ is the Plummer core radius,
and the index $\alpha$ governs the outer falloff.  Here $n_{\rm H}(r)$
is the total hydrogen-nucleus number density, defined as $n_{\rm H}(r) = n_{\rm HI}(r) + 2\,n_{\rm H_2}(r)$, where $n_{\rm HI}$ is the number density of atomic hydrogen and
$n_{\rm H_2}$ is the number density of molecular hydrogen molecules.
The factor of 2 accounts for the two hydrogen nuclei per H$_2$ molecule.
We define the radial HI abundance as
\begin{equation}
  f_{\rm HI}(r) = \frac{n_{\rm HI}(r)}{n_{\rm H}(r)}
  = \frac{n_{\rm HI}(r)}{n_{\rm HI}(r) + 2\,n_{\rm H_2}(r)}.
  \label{eq:fHI_definition}
\end{equation}
In \texttt{hinspheres}, the radial
HI abundance $f_{\rm HI}$ is 
parameterized in discrete form.

The spin temperature follows a
Plummer-like profile,
which offers the flexibility to describe a wide range of
temperature gradients, from centrally heated cores ($T_0 > T_1$)
to externally heated shells ($T_0 < T_1$):
\begin{equation}
  T(r) = T_1 + \frac{T_0 - T_1}{1 + (r / r_{\rm T})^{p}},
  \label{eq:temperature}
\end{equation}
where $T_0$ is the central temperature, $T_1$ is the ambient
temperature, $r_{\rm T}$ is the transition radius, and $p = 2$ is used by default. Throughout this paper, $T$ in temperature profiles denotes
the HI spin temperature, defined by the ratio of the hyperfine
level populations. The spin temperature $T$ and the kinetic temperature $T_{\rm kin}$ have $T=T_{\rm kin}$ under the assumption of local
thermodynamic equilibrium (LTE). 

The model includes both infall and rotational velocity fields.  The
free-fall velocity at radius $r$ is
\begin{equation}
  v_\mathrm{ff}(r) = \sqrt{\frac{2\,G\,M(r)}{r}},
  \label{eq:freefall}
\end{equation}
where $M(r)$ is the enclosed mass within radius $r$:
\begin{equation}
  M(r) = \int_0^r 4\pi r'^2\, n_{\rm H}(r')\, \mu_\mathrm{H}\, m_\mathrm{H}\, dr',
  \label{eq:enclosed_mass}
\end{equation}
with $\mu_\mathrm{H} = 1.40$ the mean gas mass per H nucleus
\citep{McKeeOstriker2007}.  The actual infall velocity is
$v_\mathrm{inf}(r) = f_\mathrm{ff} \cdot v_\mathrm{ff}(r)$, where
$f_\mathrm{ff}$ is a free parameter ($0 \leq f_\mathrm{ff} \leq 1$).  For
a line of sight at impact parameter $b$, the radial infall velocity is
projected onto the line of sight:
\begin{equation}
  v_\mathrm{los}(b, z) = v_\mathrm{inf}(r) \cdot \frac{z}{r},
  \label{eq:los_velocity}
\end{equation}
where $r = \sqrt{b^2 + z^2}$ and $z$ is the depth along the line of
sight ($z > 0$ toward the observer). Rigid-body rotation at angular velocity
$\Omega = v_\mathrm{rot} / R_\mathrm{out}$ is included as an additional
velocity component:
\begin{equation}
  v_\mathrm{rot,los} = \Omega \left(-\sin(\mathrm{PA})\,y -
  \cos(\mathrm{PA})\,x\right),
  \label{eq:rotation_los}
\end{equation}
where $\mathrm{PA}$ (position angle) is the orientation of the rotation axis
on the sky.

The peak line-center optical depth of a layer with HI column density
$N_\mathrm{HI} = n_\mathrm{HI} \cdot \delta l$, spin temperature
$T$, and velocity dispersion $\sigma_v$ is
\begin{equation}
  \tau_0 = \frac{3\,h\,c^3\,A_{10}}{32\pi\,k_\mathrm{B}\,\nu_{21}^2}
           \frac{n_\mathrm{HI} \cdot \delta l}{T\,\sigma_v\,\sqrt{2\pi}},
  \label{eq:tau0}
\end{equation}
where $A_{10} = 2.884 \times 10^{-15}\,\mathrm{s}^{-1}$ is the Einstein spontaneous transition probability for the 21\,cm transition \citep{Gould1994}, $\nu_{21} = 1.420\,\mathrm{GHz}$
is the line frequency, and the column density is in
$\mathrm{cm}^{-2}$ and $\sigma_v$ in cm/s.  The local velocity distribution is assumed to be Gaussian:
\begin{equation}
  \tau(v) = \tau_0 \exp\!\left(-\frac{(v - v_c)^2}{2\sigma_v^2}\right),
  \label{eq:tau_gaussian}
\end{equation}
where $v_c$ is the layer's center-of-mass velocity projected onto the
line of sight.  The total local velocity dispersion is the quadrature sum of
thermal and turbulent contributions:
\begin{equation}
  \sigma_v = \sqrt{\sigma_\mathrm{th}^2 + \sigma_\mathrm{turb}^2},
  \label{eq:sigma_v}
\end{equation}
where $\sigma_\mathrm{th} = \sqrt{k_\mathrm{B}\,T_\mathrm{kin} / m_\mathrm{H}}$
and $\sigma_\mathrm{turb}$ is a free parameter. 
In the dense, well-shielded interiors of molecular clouds, collisional coupling and Wouthuysen--Field pumping by Ly\(\alpha\) photons drive the HI spin temperature $T$ close to the gas kinetic temperature \(T_\mathrm{kin}\) \citep{Field1958, Shaw2017}. In the low-density envelope, spin temperature may differ from \(T_\mathrm{kin}\), but the HI column density there is small, so its contribution to the modeled spectrum is limited. Therefore, we adopt the approximation \(T_\mathrm{kin} \sim T\).

We model the observed HI spectrum as background emission transmitted
through the layered cloud.  The background brightness temperature
$T_\mathrm{bg}(v)$ enters the far side of the cloud and is attenuated
and re-emitted by each layer in turn.  Denoting the brightness
temperature after passing through layer $i$ as $T^{(i)}(v)$, the
radiative transfer is solved iteratively from the far side to the
observer:
\begin{equation}
  T^{(i)}(v) = T^{(i-1)}(v) \cdot e^{-\tau_i(v)}
              + T_i \cdot \left(1 - e^{-\tau_i(v)}\right),
  \label{eq:rt_iterative}
\end{equation}
with the initial condition $T^{(0)}(v) = T_\mathrm{bg}(v)$. The iteration proceeds from the far side of the cloud toward the observer ($i=1,\,2,\,$…$,\,N$), so that each layer receives the emergent brightness from all layers behind it. After the last layer
the emergent brightness equals the predicted observed spectrum,
which is $T_\mathrm{obs}(v)$.  The HI absorption spectrum is
defined as
\begin{equation}
  \Delta T(v) = T_\mathrm{bg}(v) - T_\mathrm{obs}(v),
  \label{eq:absorption}
\end{equation}
where positive values indicate absorption.  Because the cloud has a
temperature gradient (warm outer layers and a cold interior), the
result is not necessarily pure absorption: when the outer layers are
warmer than $T_\mathrm{bg}(v)$, they contribute emission that can
partially fill in or even reverse the absorption dip
\citep{Goldsmith2007}.

%--------------------------------------------------------------

%%======================================================================
%% 3. Data Preparation and Fitting
%%======================================================================
\section{Workflow of \texttt{hinspheres}}
\label{sec:fitting}

\subsection{Data Preparation}
\label{sec:data_prep}

The radiative-transfer fitting loop in \texttt{hinspheres} requires two observed data cubes as inputs: the HI brightness-temperature cube
$T_\mathrm{obs}(v)$ and the background HI emission cube
$T_\mathrm{bg}(v)$.  The user may supply both cubes directly (e.g.\
from an independent background extraction), or provide only the
observed data cube and use the built-in tool of \texttt{hinspheres} to
extract the background.

In the built-in background extraction tool, the background is estimated
pixel-by-pixel by fitting a high-order polynomial (default order~5) to
the line wings outside the HINSA velocity range.  The velocity mask
that defines the absorption region is selected interactively by the
user: a single pair of velocity bounds may be chosen for the entire
data cube, or, for clouds with more complex kinematics, the user may
pick multiple anchor positions on a 2D map and the program
automatically interpolates smooth per-pixel mask boundaries by distance
weighting.  In addition, the user may mark extra velocity ranges to
mask out other complex HI structures along the line of sight (e.g.\
unrelated HINSA components on the Galactic plane).  Channels falling inside the HINSA mask, together with any additional excluded spectral ranges, are omitted when performing the polynomial fit.  The resulting background estimate
$T_\mathrm{bg}(v)$ is saved as a FITS file for use by the fitting
pipeline.

\subsection{Fitting Algorithm}
\label{sec:algorithm}

In \texttt{hinspheres}, the fitting process uses the background HI emission cube as a reference. It iteratively adjusts the parameters of the spherical cloud to generate simulated data cubes, which are then fitted against the observed data cube. 
The fitting pipeline jointly determines the cloud
structure, kinematics, and $\mathrm{HI}$ abundance profile by minimizing
the discrepancy between the model and the observed data cube.  All
physical parameters are normalized to $[0, 1]$ before optimization.  This
ensures that all parameters contribute equally regardless of their
physical scales, which span several orders of magnitude.

In a 2D projected map of a spherically symmetric cloud, the pixels in
an annulus at impact parameter $b$ occupy an area proportional to
$b\,db$, so equal-per-pixel weighting would result in the outer annuli dominating the cost function.  We instead apply a radial weight $w(b) = 1/b$ to
the squared residuals, normalized by the total weight so that the cost
is independent of map size:
\begin{equation}
  S^2 = \frac{\sum_i w(b_i) \, \epsilon_i^2}{\sum_i w(b_i)},
  \label{eq:weighted_residual}
\end{equation}
where $\epsilon_i$ is the residual at pixel $i$ (model minus
observation in brightness temperature).  This gives every radial
annulus approximately equal statistical weight.  The central pixel is
assigned the weight of the innermost pixel (impact parameter set to
one pixel).  In \texttt{hinspheres} we minimize $\chi^2 = S^2 / \sigma^2$,
where $\sigma$ is the noise level of the data. The weighted residual and $\chi^2$ are computed within $r\le R_\mathrm{out}$.

The $\mathrm{HI}$ abundance profile supports two encoding modes.  In
\emph{direct} mode, each shell has an independent $f_\mathrm{HI}$ value,
yielding $N_\mathrm{shells}$ free parameters.  In \emph{parametric}
mode, the profile is unimodal: the abundance peaks at a shell with
value $f_\mathrm{HI,peak}$ and decreases monotonically via multiplicative
factors ($\leq 1$, one factor per shell boundary,
$N_\mathrm{shells}-1$ factors in total) on both sides of the peak.  This
peaked structure is physically motivated: in molecular clouds,
far-UV radiation photodissociates H$_2$ into HI at the cloud surface,
while the interior is shielded and dominated by H$_2$, producing a
radial HI abundance profile that naturally peaks between the fully
dissociated surface and the fully molecular core.  The parametric mode
yields $N_\mathrm{shells}$ continuous abundance parameters
($f_\mathrm{HI,peak}$ plus the $N_\mathrm{shells}-1$ multipliers) and
a discrete peak-shell index, trading a modest increase in parameters
for a physically regular profile.  The algorithm performs an
exhaustive grid scan, each candidate peak shell triggers an independent
optimization over the continuous parameters, and the overall best
solution is selected. 
Global optimization is performed with the Covariance Matrix Adaptation
Evolution Strategy (CMA-ES) \citep{Hansen2016}, a derivative-free method
well suited to non-convex landscapes with multiple local minima.
Convergence is assessed from the step size and function-value tolerance.

\subsection{Parameter Degeneracies}
\label{sec:degeneracy}

To illustrate the inherent parameter degeneracies of the spherical
HINSA model, we run Markov chain Monte Carlo (MCMC) on a synthetic $5\times10^6$\,yr core whose
density, temperature, and kinematic parameters are listed in
Table~\ref{tab:synth_params}.  To minimize the impact of spatial
smearing on the degeneracy geometry, we use a beam size of 
1.0\,arcmin and a pixel scale of 0.5\,arcmin. The velocity resolution is
0.2\,km/s.  Gaussian noise with a root-mean-square (RMS) of $\sigma = 0.17$\,K
per channel
(the noise level of the CRAFTS survey, \citealt{china2024}) is added to the data.

\begin{table*}[htbp]
\centering
\caption{Physical and observational parameters of the synthetic cores.}
\label{tab:synth_params}
\begin{tabular}{llc}
\toprule
Category & Parameter & Value \\
\midrule
\multirow{3}{*}{Density} & $n_0$ ($\mathrm{cm}^{-3}$) & $1.6 \times 10^{4}$ \\
& $r_0$ (pc) & 0.146 \\
& $\alpha$ & 2.0 \\
\midrule
\multirow{3}{*}{Temperature} & $T_0$ (K) & 10.0 \\
& $T_1$ (K) & 58.0 \\
& $r_{\rm T}$ (pc) & 0.73 \\
\midrule
\multirow{2}{*}{Others} & $R_\mathrm{out}$ (pc) & 1.46 \\
& $N_\mathrm{shells}$ & 7 \\
\midrule
\multirow{4}{*}{Kinematics} & $\sigma_\mathrm{turb}$ ($\mathrm{km/s}$) & 0.2 \\
& $f_\mathrm{ff}$ & 0.1 \\
& $v_\mathrm{rot}$ ($\mathrm{km/s}$) & 0.2 \\
& PA (deg) & 45.0 \\
\midrule
\multirow{4}{*}{Observation} & Distance (pc) & 300 \\
& Beam (arcmin) & 3.0 \\
& Pixel scale (arcmin) & 1.5 \\
& Velocity resolution ($\mathrm{km/s}$) & 0.2 \\
& Angular radius of $R_\mathrm{out}$ $^\dag$ (arcmin) & 16.7 \\
& Angular radius of $r_0$ $^\dag$ (arcmin) & 1.67 \\
\midrule
\multirow{1}{*}{Ages} & $t$ (yr) & $10^5,\; 5\times10^6,\; 10^7,\; 5\times10^8$ \\
\bottomrule
\end{tabular}
\vspace{2pt}

{\footnotesize $\dag$~Angular radii are computed as $\theta=R/d$ at the adopted distance of 300~pc.}
\end{table*}

\begin{figure*}[htbp]
\centering
\includegraphics[width=0.9\textwidth]{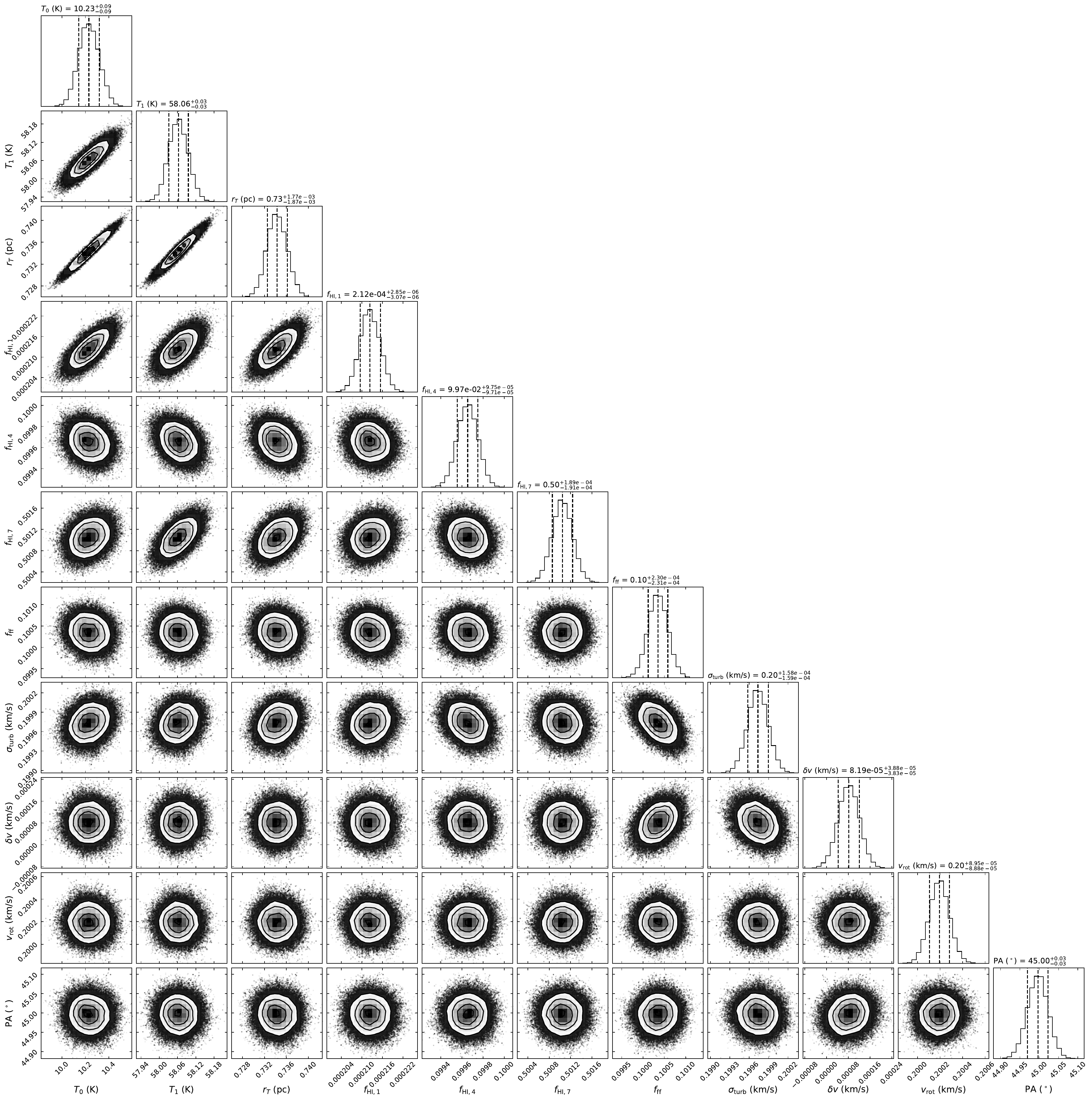}
\caption{Corner plot for the $5\times10^6$\,yr synthetic core. During the MCMC fitting, the density parameters are fixed to their true value. For clarity, only layers 1, 4, and 7 of the seven-layer HI abundance are shown.}
\label{fig:corner_part1}
\end{figure*}

We fix the density to the synthetic input values
($n_0 = 16000$~cm$^{-3}$, $r_0 = 0.146$~pc, $\alpha = 2.0$) and fit
the remaining free parameters with MCMC.  The resulting corner plot
(Figure~\ref{fig:corner_part1}) uses 48 walkers and $10^4$ iterations,
discarding a burn-in of roughly 1700 steps per walker. The Gelman--Rubin
statistic satisfies $\hat R \in [1.02,\,1.04]$ for all free parameters.
We use a seven-shell model in the fitting. For clarity, only the HI abundances of shells~1, 4,
and~7 are shown.  The three temperature parameters ($T_0$, $T_1$,
$r_{\rm T}$) are positively correlated. A higher $T_0$ or $T_1$ can be
compensated by a larger $r_{\rm T}$ to produce a similar overall temperature
structure.  Some $f_\mathrm{HI}$ parameters also correlate with
temperature, for example $f_\mathrm{HI,1}$ is positively correlated with $T_0$ and
$f_\mathrm{HI,7}$ is positively correlated  with $T_1$. It is because a joint change in temperature and HI density can maintain the same optical depth.
Apart from the correlations mentioned above, most free parameters exhibit near-circular contours in the joint probability distributions, suggesting no strong degeneracy among them. 

%%======================================================================
%% 4. Validation
%%======================================================================
\section{Validation on Synthetic Data}
\label{sec:validation}

\subsection{Synthetic cores and their fitting results}
\label{sec:synthetic}

To validate our fitting framework, we generate 7-shell synthetic
cores as in Table~\ref{tab:synth_params} and fit them with \texttt{hinspheres}. The density, temperature, and HI abundance profiles of these cores are shown in Figure~\ref{fig:radial_profiles}.
The four cores share the same density, temperature, and kinematic
parameters.  The central density $n_0$ and the total radius $R_\mathrm{out}$
are adopted from Model~1 of \citet{Goldsmith2007}, while the ambient
temperature $T_1 = 58$\,K follows the interstellar radiation
field (ISRF) model of \citet{LeBourlot1993}.  The four cores have different ages, and their
$\mathrm{HI}$ abundance follows the time-dependent chemical model of
\citet{Goldsmith2005}:
\begin{equation}
  f_\mathrm{HI}(r, t) = 1 - \frac{2k' n_{\rm H}(r)}
  {2k' n_{\rm H}(r) + \zeta_\mathrm{H_2}}
  \left[1 - \exp\!\left(-\frac{t}{\tau(r)}\right)\right],
  \label{eq:xhi_profile}
\end{equation}
where $k' = 1.2 \times 10^{-17}$\,cm$^3$\,s$^{-1}$ is the H$_2$ formation rate coefficient on grains, $\zeta_\mathrm{H_2} = 5.2 \times
10^{-17}$\,s$^{-1}$ is the cosmic-ray H$_2$ dissociation rate \citep{VanDerTak2000}, and
$\tau(r) = 1/[2k' n_{\rm H}(r) + \zeta_\mathrm{H_2}]$ is the local HI-to-H$_2$
conversion timescale.  Panel (d) of Figure~\ref{fig:radial_profiles} shows the
resulting $f_\mathrm{HI}(r)$ for four ages, produced by the four
evolutionary stages summarized in Table~\ref{tab:synth_params}.

We adopt an angular resolution of $3\arcmin$ and a velocity resolution
of $0.2$\,km/s, approximately matching the FAST beam and the CRAFTS survey
channels~\citep{Li2018, china2024}.  At the adopted distance of
$300$\,pc, the Plummer core radius $r_0 = 0.146$\,pc subtends
$1.7\arcmin$, marginally resolved by the $3\arcmin$ beam.  To mimic
realistic observing conditions, we do not use a uniform constant
background. Instead, the synthetic $\mathrm{HI}$ background is taken from
an off-position spectrum in the L1574 region~\citep{Zuo2018}, regridded
to the same spatial and velocity grid.  This single fixed background
spectrum is used for all four evolutionary ages. A CRAFTS level of noise
($\sigma_n=0.17$~K RMS per $0.2$~$\mathrm{km/s}$\ channel, \citealt{china2024}) 
is added to the synthetic cores.

\begin{figure}[htbp]
  \centering
  \includegraphics[width=\columnwidth]{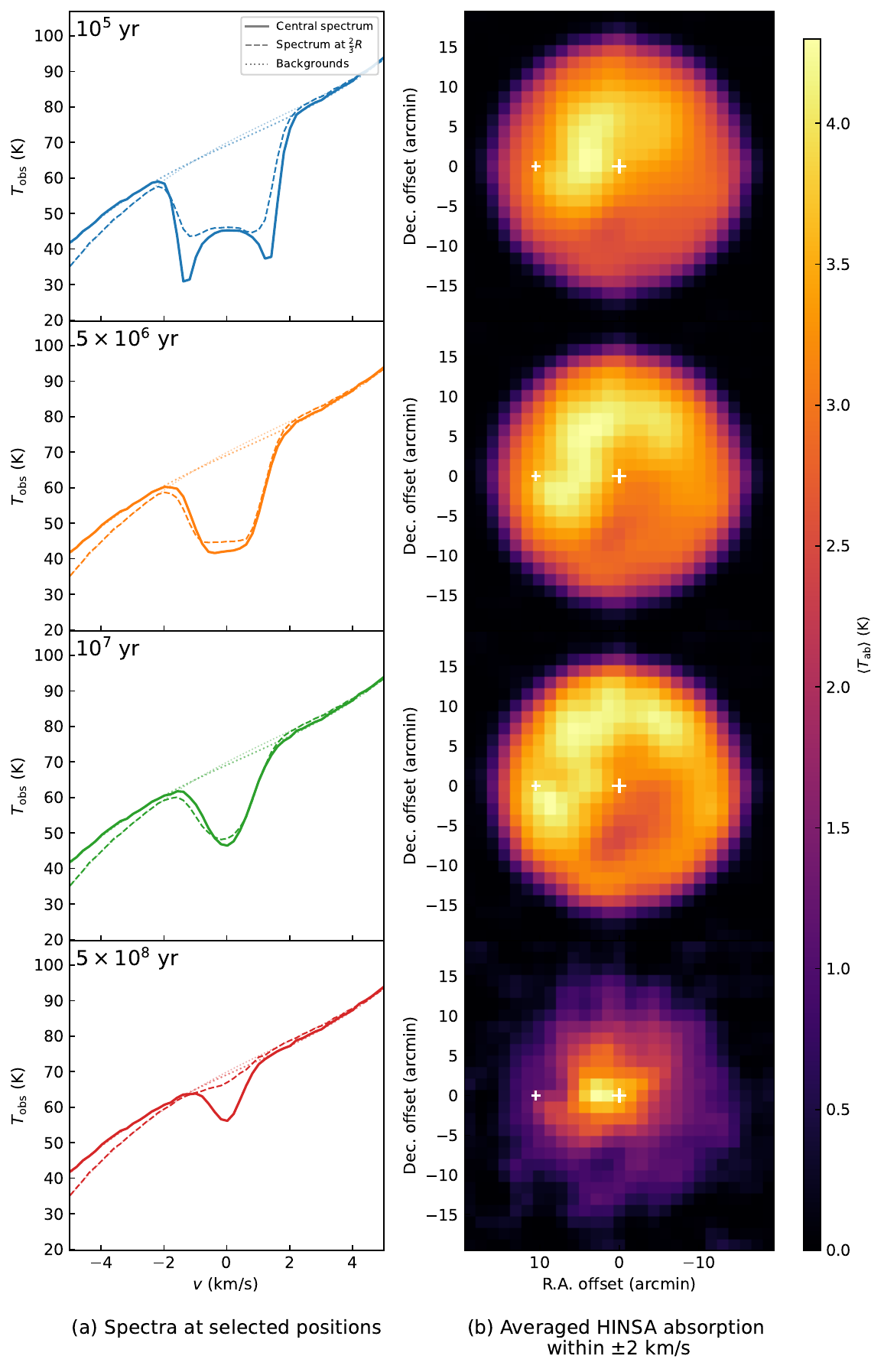}
  \caption{HINSA spectra and averaged absorption maps of the four synthetic cores.  
(a) HINSA spectra at the center (solid) and at $\frac{2}{3}\,R_\mathrm{out}$ on the eastern side (dashed) of the simulated cores. The corresponding HINSA-free background spectra at these two positions are shown as dotted lines.  
(b) Averaged absorption over $\pm2$ km/s for each synthetic core. White `+' symbols mark the center and the $\frac{2}{3}\,R_\mathrm{out}$ positions.}
  \label{fig:validation_grid}
\end{figure}

Figure~\ref{fig:validation_grid} shows the spatial and spectral
structure of the synthetic cores. The synthetic HI spectra vary strongly with age. At the youngest
stage ($10^5$~yr) the HI abundance is high throughout the core,
producing a deep absorption line whose central optical depth is strongly
saturated, and the projected infall velocity gradient along the line of
sight gives the profile a double-peaked, non-Gaussian shape.  At
$5\times10^6$~yr the absorption remains non-Gaussian but weaker, and by
$10^7$ and $5\times10^8$~yr the declining HI abundance brings the line
shape close to a single Gaussian at the CRAFTS noise level. As evolution proceeds, the HINSA absorption depth decreases, making the core appear progressively fainter in the averaged absorption map.

The degeneracy between \(n_\mathrm{HI}\) and the HI abundance \(f_\mathrm{HI}\) arises because the HINSA optical depth only constrains the product \(n_\mathrm{HI} \cdot \delta l / T\) (Equation~\ref{eq:tau0}). An independent measurement of the total hydrogen column density \(N_\mathrm{H}\), for instance, from dust continuum or extinction data, breaks this degeneracy by fixing the overall density normalization. To more realistically assess the potential errors of \texttt{hinspheres} in practical applications, we construct a synthetic \(N_\mathrm{H}\) map from the true density profile, add Gaussian noise mimicking the Planck 353~GHz thermal dust optical depth \(\tau_{353}\) observations \citep{Planck2013}, and then fit the Plummer parameters from the noisy map. The fitted values of \(n_0\), \(r_0\), and \(\alpha\) are subsequently fixed in the \texttt{hinspheres} fitting. Specifically, the total hydrogen column is \(N_\mathrm{H} = \tau_{353}/\sigma_{353}\), with \(\sigma_{353} = 6.3\times10^{-27}\)~cm\(^2\) being the dust absorption cross‑section per H nucleon at 353~GHz for the diffuse ISM (\citealt{Planck2013}, Fig.~21). We generate a synthetic \(N_\mathrm{H}\) map (\(20\times20\) pixels, \(3\times3\)~pc) by projecting the Plummer density profile, convolving it to the Planck \(5\arcmin\) beam, and adding Gaussian noise with per‑pixel \(\sigma_{\tau_{353}} \sim 2.5\times10^{-6}\), a typical value for diffuse high‑latitude ISM from the Planck modified blackbody fit uncertainty maps \citep{Planck2013}. This corresponds to \(\sigma_{N_\mathrm{H}} = \sigma_{\tau_{353}}/\sigma_{353} \sim 4\times10^{20}\)~cm\(^{-2}\). The obtained noisy map is shown in panel (c) of Figure~\ref{fig:radial_profiles}. As in Table~\ref{tab:nh_recovery}, fitting this noisy map recovers all three Plummer parameters within 10\% errors. The resulting density parameters are then used as fixed inputs in the subsequent \texttt{hinspheres} fitting.

\begin{table}[htbp]
\centering
\caption{Recovery of Plummer density parameters from the synthetic
  $N_\mathrm{H}$ map with Planck-like noise.}
\label{tab:nh_recovery}
\begin{tabular}{lcccc}
\toprule
Parameter & Input & Fitted & 68\% CI & Error \\
\midrule
$n_0$ ($10^4$\,$\mathrm{cm}^{-3}$) & 1.60 & 1.74 & $[1.44,\,2.45]$ & +9\% \\
$r_0$ (pc)               & 0.146 & 0.143 & $[0.112,\,0.164]$ & $-$2\% \\
$\alpha$                 & 2.00 & 2.08 & $[2.00,\,2.14]$ & +4\% \\
\bottomrule
\end{tabular}
\end{table}

With the density parameters fixed, we fit the remaining 15
parameters, including three temperature parameters, peak HI abundance
$f_\mathrm{HI,peak}$, six abundance multipliers, infall fraction
$f_\mathrm{ff}$, turbulent velocity $\sigma_\mathrm{turb}$, velocity
offset $\delta v$, rotation velocity $v_\mathrm{rot}$, and rotation
position angle PA, together with the discrete peak-shell index
$k_\mathrm{peak}$ (grid-scanned, see Section~\ref{sec:algorithm}). We use 50 candidate solutions per generation and at most $10^5$ generations for the CMA-ES fitting in \texttt{hinspheres}.
Recovery errors for the abundance profile are
reported as the unweighted mean over the seven shells of the
per-shell relative errors,
$\langle\delta f_\mathrm{HI}\rangle =
\frac{1}{7}\sum_k |f_\mathrm{HI,k}^\mathrm{fit}
- f_\mathrm{HI,k}^\mathrm{true}| / f_\mathrm{HI,k}^\mathrm{true}$.
The all-parameter diagnostic $\langle\delta_\mathrm{all}\rangle$
extends this to the mean relative error over the seven shell
abundances together with the eight scalar temperature and kinematic
parameters ($T_0$, $T_1$, $r_{\rm T}$, $\sigma_\mathrm{turb}$, $\delta v$,
$f_\mathrm{ff}$, $v_\mathrm{rot}$, PA).

Table~\ref{tab:synrec} summarizes the fitting results.  With the opacity calibration and the beam
convolution included self-consistently in the forward model, the
abundance profile is recovered without pathological shell errors at
any stage. The all-parameter diagnostic ranges over
$\langle\delta_{\mathrm{all}}\rangle = 4.6$--$6.8\%$ across the four
stages, lowest for the mid-age stages ($4.6\%$ and $4.9\%$ at
$5\times10^6$ and $10^7$\,yr) and slightly higher for the young
($6.8\%$) and oldest ($6.3\%$) stages.  The shell-averaged abundance
errors grow with age, $\langle\delta f_{\mathrm{HI}}\rangle = 4.1\%$
($10^5$\,yr) rising to $10.4\%$ ($5\times10^8$\,yr), with the largest
per-shell errors in the outer shells of the older stages (up to
$\sim 18\%$ at $5\times10^8$\,yr).  For the $10^5$\,yr stage the
innermost shell is recovered to within $5.4\%$ (fitted $42.4\%$
against the true $44.8\%$), the largest relative error being $13.6\%$
in shell~3. The temperature parameters are also recovered well. The $T_0$ errors are $38.8\%$, $7.0\%$, $7.9\%$, and $2.6\%$ for the four cores, respectively.
 $r_{\rm T}$ is recovered to within $4\%$
except for the youngest stage ($13.3\%$), which is because of the strong optical-depth saturation in the young core center.  $T_1$ is recovered to within
$3.4\%$ in all cases. The infall fraction $f_\mathrm{ff}$,
turbulent velocity $\sigma_\mathrm{turb}$, and rotation parameters are
reliably recovered across all stages ($\lesssim14\%$). 
At the three older stages, the best-fit residual is $0.0285$,
consistent with the RMS of $\sim 0.17$~K of our data.  The $10^5$~yr stage shows a higher residual
($0.0784$, RMS $\sim 0.28$~K). It is the only stage whose residual
substantially exceeds the noise level. 

\input{synrec_table.tex}

%--------------------------------------------------------------

\subsection{Combined Resolution--Noise Analysis}
\label{sec:resolution_noise}

To assess how observational conditions affect parameter recovery, we perform a joint two‑dimensional sweep over beam size and noise level with the density parameters fixed. We use the \(5\times10^6\)~yr synthetic core and set \(n_0\), \(r_0\), and \(\alpha\) to the values derived in Section~\ref{sec:synthetic}. All remaining 15 continuous parameters are left free. We sweep seven beam sizes (\(1\arcmin\), \(3\arcmin\), \(6\arcmin\), \(10\arcmin\), \(15\arcmin\), \(20\arcmin\), \(30\arcmin\)) and seven noise levels (\(0.0\), \(0.05\), \(0.10\), \(0.17\), \(0.30\), \(0.60\), \(1.0\)~K RMS). Since the synthetic cloud is placed at 300~pc, these beam sizes are equivalent to observing the same cloud at distances from 0.1 to 3~kpc with a fixed $3\arcmin$ beam. At each beam size, the data are regridded to a pixel scale of half the beam full width at half maximum (FWHM), and the velocity channels are interpolated to \(0.2\)~\(\mathrm{km/s}\). Each combination of beam and noise is run with 20 different random seeds, giving \(7\times7\times20=980\) total fits.

\begin{figure}[htbp]
\centering
\includegraphics[width=\columnwidth]{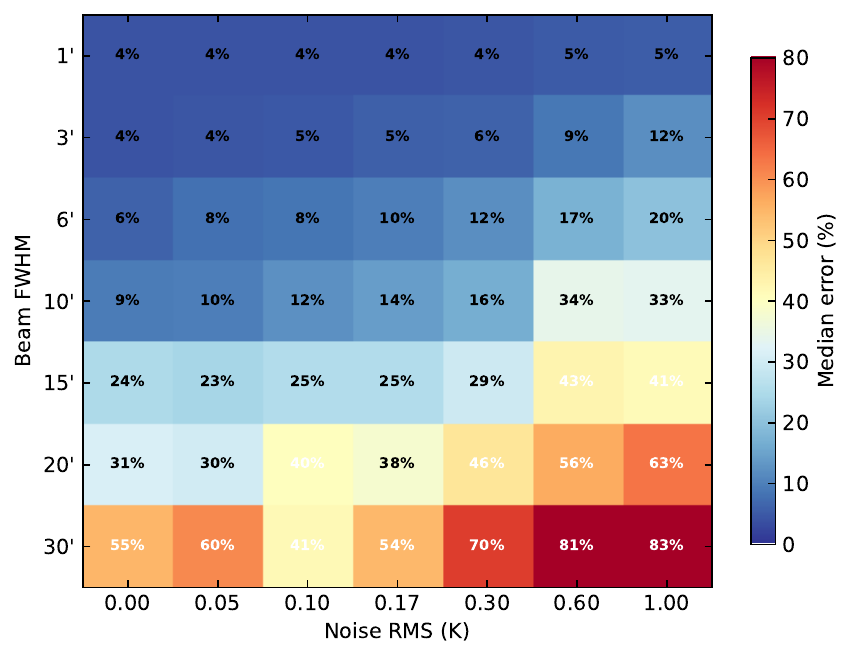}
\caption{Median relative recovery error $\langle\delta_{\rm all}\rangle$ averaged over all 15 fitted parameters, as a function of beam FWHM and noise RMS. 
  The result is obtained from the $5\times10^6$\,yr synthetic core with the density profile fixed to the
  $N_\mathrm{H}$-map derived values. 20 random noise realizations are used in the calculation.}
\label{fig:resolution_noise}
\end{figure}

\begin{figure*}[htbp]
\centering
\includegraphics[width=\textwidth]{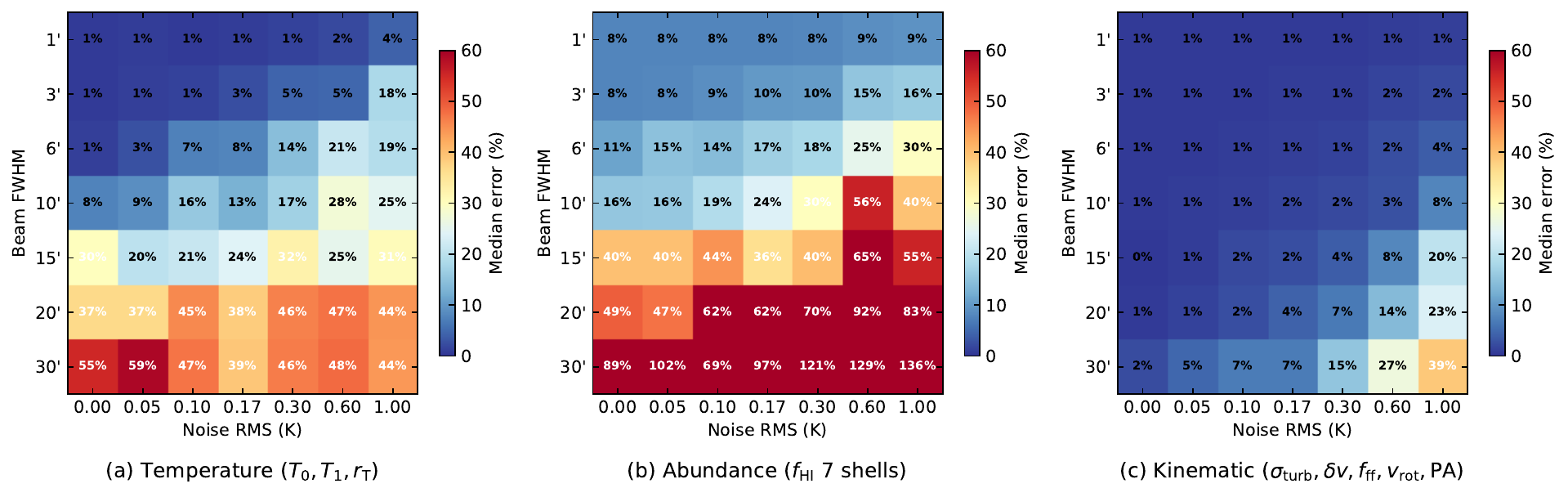}
\caption{Same as Figure~\ref{fig:resolution_noise} but grouped by
  parameter family: (a) temperature structure ($T_0$, $T_1$,
  $r_{\rm T}$), (b) HI abundance (seven shells), (c) kinematic
  parameters ($\sigma_\mathrm{turb}$, $\delta v$, $f_\mathrm{ff}$,
  $v_\mathrm{rot}$, PA).}
\label{fig:resolution_noise_grouped}
\end{figure*}

Figure~\ref{fig:resolution_noise} shows the median relative error over all 15 fitted parameters
across the 20 seeds. In Figure~\ref{fig:resolution_noise_grouped}, we show the median relative errors separately for the three temperature parameters, the seven HI abundance layers, and the other parameters (collectively referred to as kinematic parameters).
Overall, the parameters recovered by \texttt{hinspheres} show relatively low sensitivity to beam size and noise level. For a typical FAST angular resolution and a core with \(R_{\rm out}=1.46\,\mathrm{pc}\) located within 1 kpc, the median error over all 15 parameters remains below 34\%. For common HINSA targets with RMS noise below 0.2 K, the median error is expected to be less than 16\%. The median relative error exceeds 50\% only for beams of $20\arcmin$ at noise levels of 0.6 K and above, or at $30\arcmin$ nearly independently of the noise level (equivalent to observing the same core at distances beyond 2 kpc with FAST).
With the density fixed, the temperature structure is tightly constrained at high resolution: the median error is $\sim$1\% at $1\arcmin$ and stays below 8\% up to $6\arcmin$ and below 13\% up to $10\arcmin$ at CRAFTS noise, rising to 20–59\% at $15$--$30\arcmin$. The kinematic parameters are less affected by noise and resolution. They are recovered to 1–2\% for beams up to $10\arcmin$ at CRAFTS noise, and degrade to 20–39\% at $15$--$30\arcmin$ under 1 K noise.
The shell-averaged abundance error \(\langle\delta f_\mathrm{HI}\rangle\) is most influenced by noise and beam sizes. Its median error is 8–9\% at the fully resolved $1\arcmin$ beam and grows with beam size, reaching 69–136\% at $30\arcmin$ depending on the noise level. Higher noise causes earlier degradation. At 0.6–1 K the error already reaches 25–30\% at $6\arcmin$ and rises to 40–136\% at $10$--$30\arcmin$. Even at $1\arcmin$, the error does not drop to zero but saturates at about 8\%. This floor stems from the density–abundance degeneracy: because the density profile is fixed to the $N_\mathrm{H}$-map derived values (which deviate slightly from the true profile), the fit compensates for the incorrect column by adjusting \(f_\mathrm{HI,peak}\) and the shell multipliers. This compensation is bounded by the constraints of the fitted spectrum and cannot be fully removed.

%--------------------------------------------------------------

\subsection{Effects of Density and Temperature Constraints}
\label{sec:density_T0_grid}

As discussed in the subsections above, the strong degeneracy between total hydrogen density and HI abundance in the fitting of HI radiative transfer necessitates external constraints on the density. Moreover, the inherent degeneracy between temperature and density suggests that imposing a temperature prior may also improve the fit. In the previous analysis, we broke the density–abundance degeneracy by fixing the density parameters to values derived from the total hydrogen column density map, which successfully stabilized the \texttt{hinspheres} fitting but inevitably introduced systematic errors. In this subsection, we use the $5\times10^6$~yr synthetic core in Table~\ref{tab:synth_params} to systematically investigate how different levels of constraints on the density and temperature parameters affect the \texttt{hinspheres} recovery.

We consider three density constraint schemes: (i) unconstrained; (ii) $1\sigma$ constrained, where the three density parameters $n_0$, $r_0$, and $\alpha$ are restricted to the $\pm1\sigma$ intervals obtained from the $N_\mathrm{H}$ map fit; and (iii) fixed, where $n_0$, $r_0$, and $\alpha$ are set to the best-fit values from the $N_\mathrm{H}$ map fit (identical to the approach used in the previous subsections). For temperature, we test two configurations: (i) unconstrained; and (ii) fixed $T_0$, with $T_0$ held at its true value of $10$~K for the synthetic core. In real observations, such an external constraint on $T_0$ could be provided by independent chemical tracers. For example by assuming that the excitation temperature of $^{13}$CO or C$^{18}$O is representative of the core center. We thus evaluate a total of $2\times3 = 6$ scenarios, each scenario with 10 random seeds of noise realization, and summarize the results in Table~\ref{tab:density_T0_grid}. 

The strength of the density constraint is the dominant factor governing the recovery of all physical parameters. Without any constraint, the errors on $n_0$, $r_0$, and $\alpha$ are as large as $96\%$, $409\%$, and $69\%$, respectively, and the fit wanders wildly across the broad priors. Introducing the loose $\pm1\sigma$ constraint reduces these errors dramatically to $18\%$, $13\%$, and $4.4\%$, while simultaneously the free-fall coefficient $f_\mathrm{ff}$ drops from $40.6\%$ to $11.3\%$, and the shell-averaged abundance deviation $\langle\delta f_\mathrm{HI}\rangle$ plummets from $182\%$ to $21.9\%$. This confirms that even approximate column-density priors effectively break the density–abundance degeneracy. Fixing the density exactly further reduces the abundance deviation from $21.9\%$ to $9.0\%$. 

The temperature constraint provides a secondary but quantifiable improvement, particularly when the density prior is only loose. Fixing $T_0$ to its true value, in the $\pm1\sigma$ density-constrained case, reduces the $n_0$ error from $18\%$ to $8.1\%$, $\alpha$ from $4.4\%$ to $3.0\%$, $f_\mathrm{ff}$ from $11.3\%$ to $9.5\%$, and $\langle\delta f_\mathrm{HI}\rangle$ from $21.9\%$ to $15.7\%$. When the density is already fixed exactly, the additional improvement from fixing $T_0$ is much smaller (e.g., the abundance deviation only drops from $9.0\%$ to $8.4\%$), implying that the temperature prior is most valuable when the density prior is not exact. When $T_0$ is left free, it is recovered to within $\sim6\%$ as long as the density is constrained (compared to $16.6\%$ without a density constraint), while the outer temperature $T_1$ is recovered to within $0.5\%$ in all cases (and to $0.1$–$0.2\%$ when $T_0$ is fixed). This shows that the absorption spectrum alone provides substantial constraints on the temperature structure. The residuals remain constant at $0.0288$–$0.0289$ across all scenarios, equal to the square of the $0.17$~K RMS noise, confirming that the forward model compensates for an incorrect density by adjusting the abundance profile.  

In practical applications of \texttt{hinspheres}, we therefore recommend that users prioritize applying density constraints. Even an approximate column-density prior derived from dust continuum or far-infrared emission (e.g., the $\pm1\sigma$ intervals used here) suffices to reduce the recovery errors in density parameters and abundance profiles from several hundred percent to about twenty percent, representing the most critical step in breaking the density–abundance degeneracy. If a reliable independent density estimate is available, fixing the density value is also a viable option. Regarding temperature priors, constraining $T_0$ via excitation temperatures from chemical tracers can provide a secondary improvement when the density constraint is loose.

\begin{table*}[htbp]
\centering
\caption{Recovery errors for the $5\times10^6$\,yr core under different density
  and temperature constraints.  Errors are median across 10 random
  noise realizations.  $\langle\delta f_\mathrm{HI}\rangle$
  is the unweighted mean of per-shell $f_{\rm HI}$ relative errors.}
\label{tab:density_T0_grid}
\footnotesize
\setlength{\tabcolsep}{3pt}
\begin{tabular}{lcccccc}
\toprule
& \multicolumn{3}{c}{$T_0$ free} & \multicolumn{3}{c}{$T_0$ fixed to $10$~K} \\
\cmidrule(lr){2-4} \cmidrule(lr){5-7}
Density constraint
  & None & $\pm1\sigma$ & Fixed
  & None & $\pm1\sigma$ & Fixed \\
\midrule
$n_0$ error     & $95.8\%$ & $18.0\%$ &  $9.0\%$ & $72.0\%$ &  $8.1\%$ &  $9.0\%$ \\
$r_0$ error     & $409.0\%$  & $13.4\%$ &  $2.3\%$ & $96.7\%$ & $10.7\%$ &  $2.3\%$ \\
$\alpha$ error  & $69.2\%$ &  $4.4\%$ &  $4.0\%$ & $53.9\%$ &  $3.0\%$ &  $4.0\%$ \\
$T_0$ error     & $16.6\%$ &  $5.4\%$ &  $6.2\%$ & --- & --- & --- \\
$T_1$ error     &  $0.5\%$ &  $0.2\%$ &  $0.3\%$ &  $0.2\%$ &  $0.1\%$ &  $0.1\%$ \\
$f_\mathrm{ff}$ error & $40.6\%$ & $11.3\%$ &  $3.6\%$ & $49.5\%$ &  $9.5\%$ &  $3.2\%$ \\
$\sigma_\mathrm{turb}$ error &  $0.3\%$ &  $0.3\%$ &  $0.3\%$ &  $0.3\%$ &  $0.3\%$ &  $0.3\%$ \\
$\langle\delta f_\mathrm{HI}\rangle$ & $182.0\%$ & $21.9\%$ &  $9.0\%$ & $66.4\%$ & $15.7\%$ &  $8.4\%$ \\
\midrule
Residual & 0.0288 & 0.0288 & 0.0289 & 0.0288 & 0.0288 & 0.0288 \\
\bottomrule
\end{tabular}
\end{table*}

%--------------------------------------------------------------
%%--------------------------------------------------------------------
%%Effect of $R_\mathrm{out}$ Misestimation
%%--------------------------------------------------------------------
\subsection{Effect of $R_\mathrm{out}$ Misestimation}
\label{sec:rout_misest}

The total radius $R_\mathrm{out}$ of the core can be hard to constrain
in observations, yet it affects the radiative-transfer calculation by setting the shell
boundaries. To quantify the impact of an incorrect estimate \(R_\mathrm{est}\), we run \texttt{hinspheres} on the \(5\times10^6\,\mathrm{yr}\) synthetic core (true \(R_\mathrm{out}=1.46\,\mathrm{pc}\)), sampling \(R_\mathrm{est}\) from \(0.50\,R_\mathrm{out}\) to \(1.71\,R_\mathrm{out}\). The density profile is fixed to the $N_\mathrm{H}$-map derived values from Section~\ref{sec:synthetic} (\(n_0 = 1.74\times10^{4}\,\mathrm{cm^{-3}}\), \(r_0 = 0.143\,\mathrm{pc}\), \(\alpha = 2.08\)). For each trial, we generate 10 independent noise realizations and perform a full CMA-ES optimization.

Table~\ref{tab:rout_scan} summarizes the median relative errors and \(1\sigma\) spreads across seeds for each trial \(R_\mathrm{est}\). With the correct \(R_\mathrm{est}\) (\(1.46\,\mathrm{pc}\)), all parameters recover to within \(\sim5\%\) of truth, matching the baseline results. Underestimating \(R_\mathrm{est}\) by more than \(\sim18\%\) (\(\le1.20\,\mathrm{pc}\)) raises the residual within the assumed radius from \(0.029\) to \(1.38\)–\(5.81\) and produces qualitatively incorrect fits: the errors in $T_0$, $r_{\rm T}$, and $f_{\rm ff}$ all exceed \(30\%\). Overestimating by \(23\)–\(71\%\) (\(1.80\)–\(2.50\,\mathrm{pc}\)) yields moderate residuals (\(0.11\)–\(0.27\)), but the retrieved parameters are also systematically biased, especially \(T_0\), which is overestimated by \(23\)–\(62\%\).

\begin{table*}[t]
\centering
\caption{Parameter recovery under $R_\mathrm{est}$ misestimation. Errors are median $\pm 1\sigma$ across 10 realizations.}
\label{tab:rout_scan}
\footnotesize
\setlength{\tabcolsep}{3pt}
\begin{tabular}{lcccccccc}
\toprule
$R_\mathrm{est}$ (pc) & 0.73 & 1.00 & 1.20 & \textbf{1.46} & 1.80 & 2.00 & 2.50 \\
\midrule
$T_0$ error
  & $133.3\%\pm4.4\%$ & $76.4\%\pm39.4\%$ & $111.1\%\pm104.7\%$
  & \textbf{$4.9\%\pm4.8\%$}
  & $62.1\%\pm16.8\%$ & $49.5\%\pm1.8\%$ & $23.3\%\pm10.1\%$ \\
$T_1$ error
  & $82.8\%\pm0.0\%$ & $33.2\%\pm0.0\%$ & $29.6\%\pm0.0\%$
  & \textbf{$0.8\%\pm0.6\%$}
  & $4.9\%\pm4.6\%$ & $1.9\%\pm0.3\%$ & $4.6\%\pm0.9\%$ \\
$r_{\rm T}$ error
  & $62.3\%\pm4.0\%$ & $98.6\%\pm0.0\%$ & $97.7\%\pm2.0\%$
  & \textbf{$2.1\%\pm3.0\%$}
  & $23.9\%\pm9.3\%$ & $11.4\%\pm0.7\%$ & $11.9\%\pm3.7\%$ \\
$f_\mathrm{ff}$ error
  & $262.1\%\pm4.3\%$ & $128.4\%\pm4.5\%$ & $99.6\%\pm2.7\%$
  & \textbf{$5.3\%\pm2.0\%$}
  & $2.2\%\pm9.7\%$ & $1.0\%\pm1.4\%$ & $3.0\%\pm2.0\%$ \\
Residual
  & $5.8133\pm0.0214$ & $2.4612\pm0.0122$ & $1.3793\pm0.0058$
  & \textbf{$0.0288\pm0.0006$}
  & $0.2713\pm0.1027$ & $0.1086\pm0.0014$ & $0.1135\pm0.0013$ \\
\bottomrule
\end{tabular}
\end{table*}

\begin{figure}[t]
\centering
\includegraphics[width=\columnwidth]{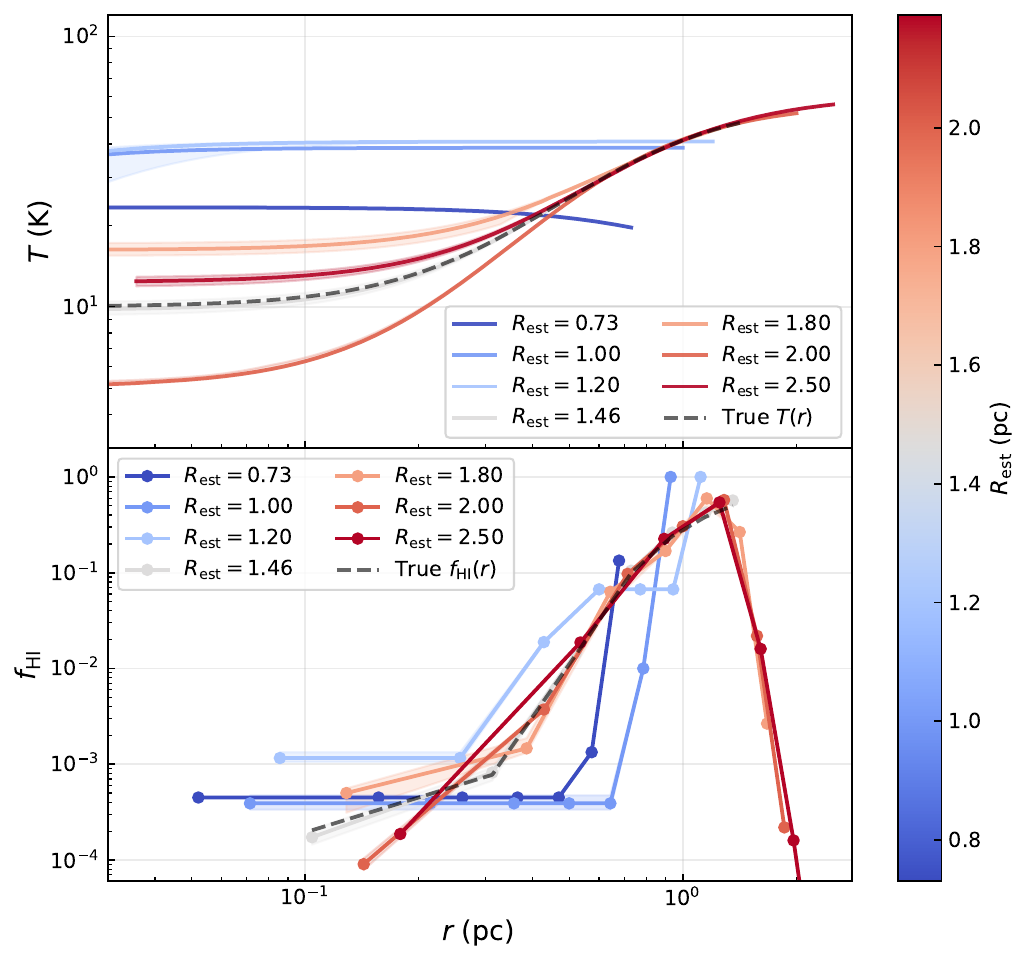}
\caption{Temperature and HI abundance profiles
  recovered under different $R_\mathrm{est}$ assumptions, compared to the true profiles (dashed black curves).  Each profile is the median across 10 noise realizations.}
\label{fig:rout_profiles}
\end{figure}

Figure~\ref{fig:rout_profiles} compares the recovered temperature (top) and HI abundance (bottom) profiles against the true ones. The true abundance increases from shell 1 (\(f_\mathrm{HI}=2\times10^{-4}\)) to shell 7 (\(f_\mathrm{HI}=0.50\)), as expected for a \(5\times10^6\,\mathrm{yr}\) core with our Plummer-like density profile. When \(R_\mathrm{est}\) is underestimated (\(\le1.20\,\mathrm{pc}\)), the fitter pushes the abundance peak to the outermost available shell, concentrating essentially all HI at the cloud edge while the inner shells become nearly empty (\(f_\mathrm{HI}<0.001\)). Concurrently, the inner temperature is overestimated by a factor of 2–4 to compensate for the missing absorption, effectively erasing the expected central-cold, outer-warm temperature gradient. With the correct \(R_\mathrm{est}\) (\(1.46\,\mathrm{pc}\)), both profiles closely track the true curves. Overestimation (\(\ge1.80\,\mathrm{pc}\)) shifts the abundance peak inward to the shell near \(1.46\,\mathrm{pc}\). The fitted profile matches the true one for \(r<1.46\,\mathrm{pc}\) but drops sharply beyond, creating an additional low-\(f_\mathrm{HI}\) envelope outside that region.

The above results expose a clear asymmetry. Underestimating \(R_\mathrm{est}\) severely compresses the available volume, causing the temperature gradient within the true cloud radius to be flattened and forcing all HI into the outermost shell. It destroys the physical temperature structure that the HINSA diagnostic relies on. In contrast, overestimating \(R_\mathrm{est}\) preserves the inner temperature gradient to within a smaller error, and the extra shells beyond \(1.46\,\mathrm{pc}\) contribute limited absorption, so the internal solution remains largely intact. Therefore, in practice, when \(R_\mathrm{out}\) is uncertain, we recommend that users adopt a generous upper bound rather than a tight lower bound, as underestimating the outer radius is far more damaging than overestimating it.

%%======================================================================
%% 5. Application
%%======================================================================
\section{Application to Observational Data}
\label{sec:application}

We apply the \texttt{hinspheres} pipeline to two dark cores observed in
HINSA: L1574 and L1544.  We selected these two sources because their HI centers coincide with their dust centers, their HINSA structures have been carefully studied in previous works \citep{Goldsmith2005,Zuo2018}, while no ring-like HI abundance distribution analogous to that of B227 was identified in either of them.

L1574 (or L1574b) is a low-mass dark core located at a distance of $d=300$\,pc \citep{Kawamura1998}. It exhibits clear HINSA features at around $0$\,km/s \citep{Goldsmith2005}, who found that its spectral profiles are highly consistent with those of the $^{13}$CO, C$^{18}$O, and OH molecular emission in both velocity and line width. This consistency suggests that cold HI is well mixed with the molecular gas within the core. They obtained an observed HI abundance of approximately $10^{-3}$, implying a core age on the order of $10^{6.5}$ to $10^7$ years. The core was mapped with the Arecibo 305\,m telescope's ALFA feed array as part of the observations by \citet{Zuo2018}. Although it is located close to B227 on the plane of the sky, \citet{Zuo2018} did not detect the central HI abundance depression seen in B227. Here, we utilize their destriped three-dimensional HI data cube (beam size $3.83\arcmin$, velocity resolution $0.184$\,km/s). The per-channel noise we measured from line-free channels is $\sigma \sim 50$\,mK.

L1544 is a well-studied prestellar core located at $d = 170$\,pc \citep{Galli2019}. The core exhibits extremely low gas temperatures in its inner regions, with interferometric NH$_3$ observations resolving a gas temperature drop from 12\,K to 5.5\,K within the innermost $\sim$5000\,AU at the center \citep{Crapsi2007}.
L1544 possesses an exceptionally rich molecular line spectrum. CO isotopologues are heavily depleted toward the center, with a depletion factor of $\sim$10 at the dust peak \citep{Caselli1999}. In contrast, species such as N$_2$H$^+$ and N$_2$D$^+$ remain in the gas phase throughout the core, exhibiting a high deuterium fraction \citep{Crapsi2005}. L1544 hosts a variety of complex organic molecules \citep{Vastel2014,JS2016}. Additionally, carbon-chain molecules and cyanopolyynes have been detected, revealing the chemical complexity of the source \citep{Snell1981,Bianchi2023}.
The chemical age of L1544 has been estimated using various molecular tracers and modeling approaches, yielding a wide range of values. Early estimates based on CO depletion suggested a young age of $\sim$10$^4$\,yr \citep{Caselli1999}. \citet{Vasyunin2017} employed a multilayer gas-grain chemical model and found that the peak abundances of COMs in L1544 are reached at timescales of $(1-2)\times10^5$\,yr, corresponding to the epoch when the modeled CO depletion factor matches the observed value. In contrast, sulfur chemistry modeling by \citet{Vastel2018} yielded a best-fit age of $10^6$--$10^7$\,yr.
\citet{Goldsmith2005} carried out a multi-species study of L1544, combining HINSA, OH, \(^{13}\)CO, C\(^{18}\)O, and CI observations. The spatial distributions of HINSA, \(^{13}\)CO, and C\(^{18}\)O are found closely correlated. Using a triaxial Gaussian model, they obtained an HI abundance of \(2.3\times10^{-4}\)–\(4.8\times10^{-4}\), from which they inferred a cloud age of \(10^{6.5}\)–\(10^{7}\) yr as a lower limit. 
For L1544, we use HI data cube from the FAST CRAFTS survey. The data have an angular resolution of $2.9\arcmin$ and a velocity resolution of 0.201\,km/s. The nominal RMS noise of the CRAFTS data is expected to be 0.17 K \citep{china2024}. However, the per‑channel noise measured from line‑free channels is \(\sigma \sim 91\) mK. We therefore adopt this measured value.

\begin{figure*}[htbp]
  \centering
  \includegraphics[width=0.8\textwidth]{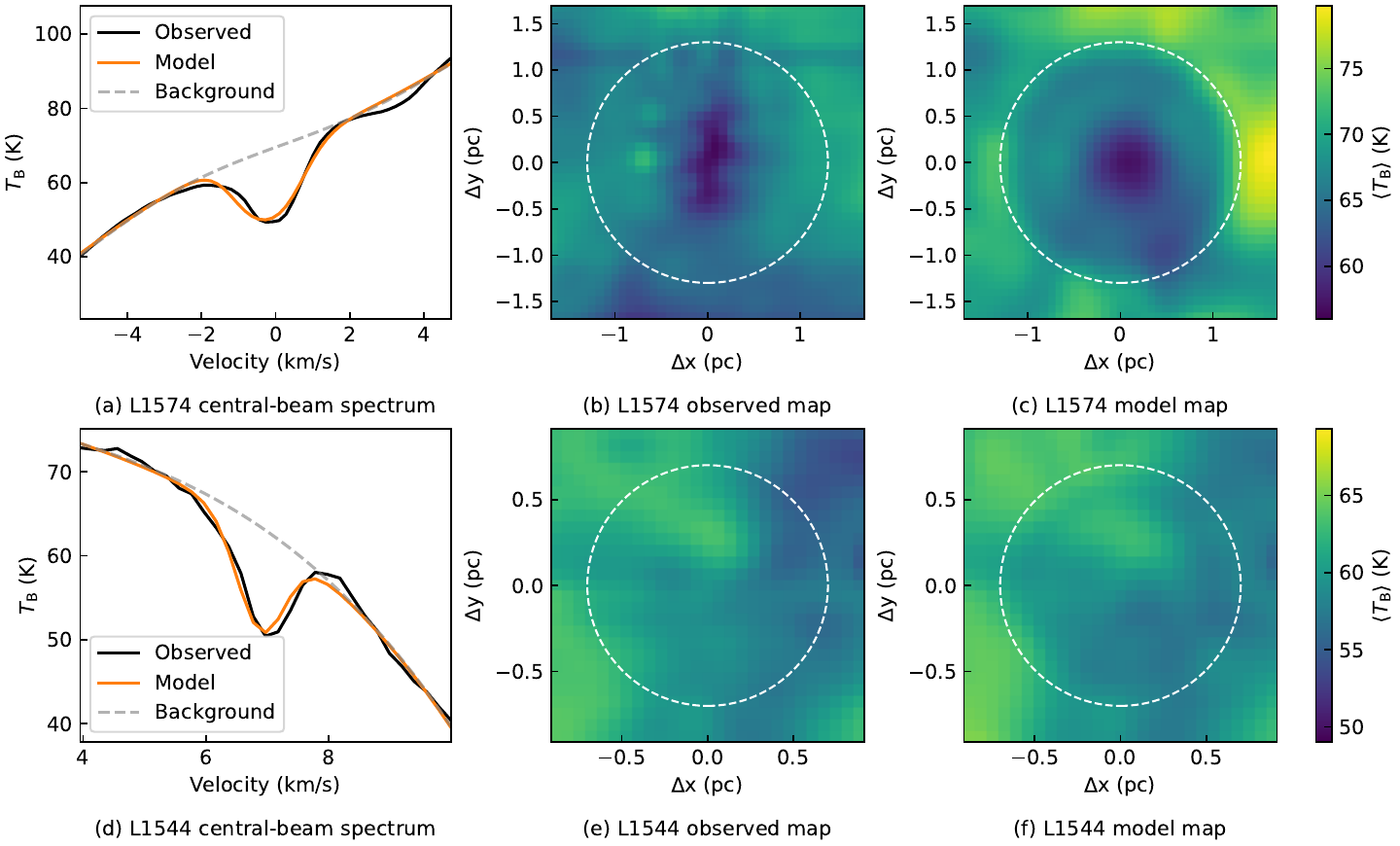}
  \caption{Central-beam spectra (left) and line-center channel-averaged maps (middle and right) for L1574 (top) and L1544 (bottom). Left: observed, best-fit model, and polynomial background spectra extracted as the Gaussian-weighted mean over the synthesized beam centered on the core ($3.83\arcmin$ for L1574 and $2.9\arcmin$ for L1544). Middle and right: observed and model cubes averaged over $\pm2$\,km/s around the fitted line-center velocity. Each map is centered on the adopted core center. The white dashed circle marks $R_{\rm out}$. }
  \label{fig:center_spec_maps}
\end{figure*}

For L1574 and L1544, we construct the total hydrogen column
density $N_\mathrm{H}$ map from the Planck 353\,GHz dust
optical depth map ($\sigma_{353}=6.3\times10^{-27}$\,cm$^2$;
\citealp{Planck2013}). We then fit the map to get the Plummer density parameters
($n_0$, $r_0$, $\alpha$; Table~\ref{tab:obs_params}). $R_{\rm out}$ and central coordinates of the two cores are determined from both the $N_\mathrm{H}$ map and the HI data. $R_{\rm out}$ are set to be 1.3 and 0.7 pc, respectively.
The background HI emission cube was extracted using the built‑in background extraction tool of \texttt{hinspheres} in its simple mode. 
For L1574, we mask -2.12--1.93\,km/s and model the background by fitting a default fifth-order polynomial over a range of $\pm 8$\,km/s around the line center. For L1544, we mask 5.24–7.35\,km/s. Because the line-of-sight velocity structure toward L1544 is complex, to avoid contamination from the component at 3.0\,km/s, we instead perform a third-order polynomial background fit over $\pm 3$\,km/s around the line center. The mean HI intensity maps and line-center spectra for both cores are shown in Figure~\ref{fig:center_spec_maps}. We then fix the density profile and optimize the remaining parameters. In the fits, we used five radial shells for both cores.

\begin{table*}[htbp]
\centering
\caption{Observational and fitted density parameters adopted for L1574 and L1544.}
\label{tab:obs_params}
\begin{tabular}{lccccccccccc}
\toprule
Core & $\alpha_\mathrm{J2000}$ & $\delta_\mathrm{J2000}$ & $d$ & $v_{\rm lsr}$ & $R_\mathrm{out}$ & Beam & $\Delta v$ & $\sigma_{\rm rms}$ & $n_0$ & $r_0$ & $\alpha$ \\
     & (h\,m\,s) & ($\degr$\,\arcmin\,\arcsec) & (pc) & (km/s) & (pc) & (\arcmin) & (km/s) & (mK) & (cm$^{-3}$) & (pc) & \\
\midrule
L1574 & 06:08:05.0 & +18\,33\,36 & 300 & 0.0 & 1.3 & 3.83 & 0.184 & $50$ & $1.60\times10^{4}$ & 0.339 & 2.63 \\
L1544 & 05:04:17.2 & +25\,12\,36 & 170 & 7.0 & 0.7 & 2.9 & 0.201 & $91$ & $1.68\times10^{5}$ & 0.078 & 2.55 \\
\bottomrule
\end{tabular}
\end{table*}

\begin{figure}[htbp]
  \centering
  \includegraphics[width=\columnwidth]{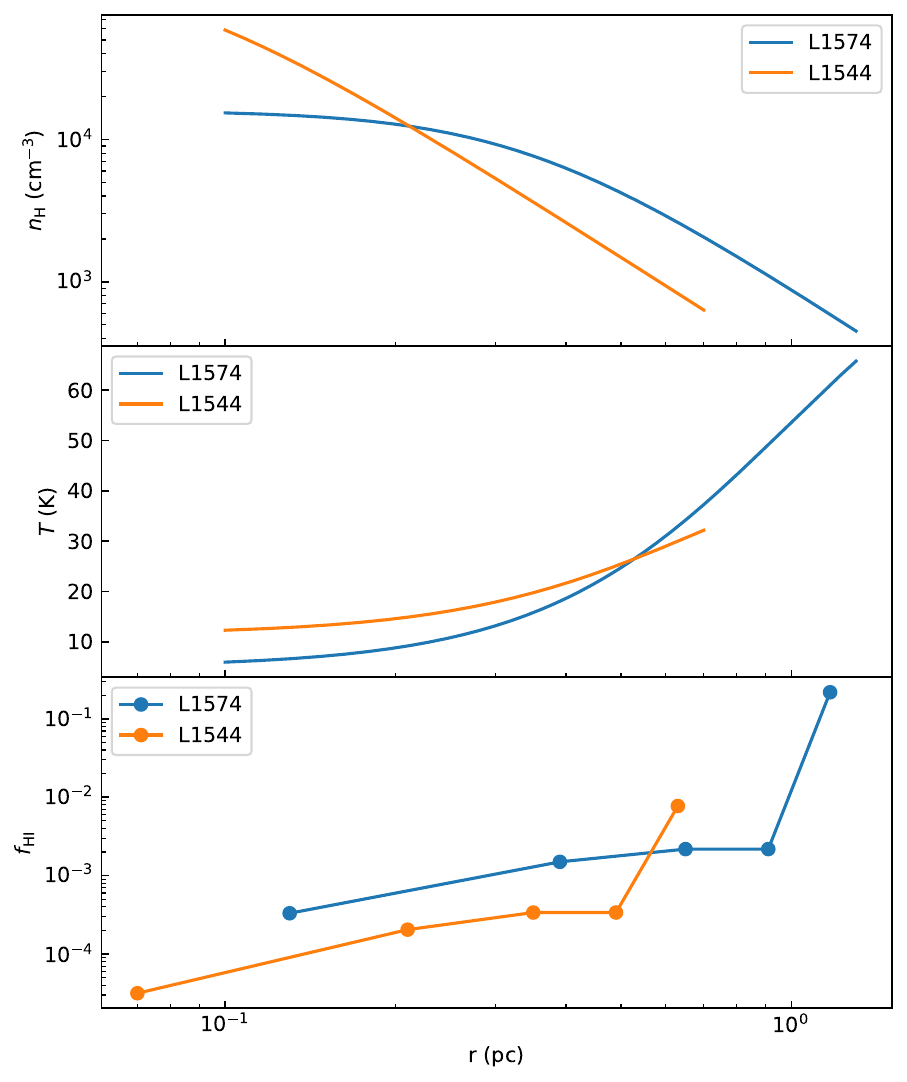}
  \caption{Radial density, temperature, and HI abundance profiles recovered for L1574 and L1544.}
  \label{fig:real_profiles}
\end{figure}

The best-fit parameters for both cores are listed in Table~\ref{tab:synrec}, and the obtained radial profiles are shown in Figure~\ref{fig:real_profiles}. 
On scales from 0.1\,pc to $R_{\rm out}$, L1544 exhibits a steeper density profile than that of L1574. Its central density, $n_0=1.68\times10^{5}$\,cm$^{-3}$ is high, while $r_0=0.078$\,pc remains below the data's resolution of 0.14\,pc. Therefore, its density profile is overall power-law-like, and $r_0$ is not sufficiently constrained. In contrast, the density profile of L1574 is closer to a Bonnor-Ebert profile, with $r_0$ close to its resolution of 0.33\,pc.
Table~\ref{tab:synrec} shows that the central temperature $T_0$ and the envelope temperature $T_1$ of L1574, and the temperature transition radius $r_{\rm T}$ of L1544, lie at the boundaries of our adopted parameter ranges ($T_0$: 5--30\,K, $T_1$: 10--100\,K, $r_{\rm T}$: 0.01\,pc to $R_{\rm out}$, marked by $\dag$). The upper bound of $r_{\rm T}$ is set to $R_{\rm out}$ so that the temperature transition occurs inside the modeled cloud. These boundary values are formal parameters of the parametric profile rather than temperature measurements, and the physical temperature structure is the resolved profile within $R_{\rm out}$ shown in Figure~\ref{fig:real_profiles}. From the resolution limit to $R_{\rm out}$, the temperatures are approximately 15.0--65.8\,K for L1574 and 13.2--32.2\,K for L1544. Both cores exhibit radial temperature profiles with lower values in central regions. Because the beam sizes of the two data sets are 3.8$\arcmin$ for L1574 and 2.9$\arcmin$ for L1544, corresponding to physical scales of 0.33 and 0.14 pc at their respective distances, the data constrain the temperature profile only on scales larger than the beam. Values at smaller radii, including the formal central temperature $T_0$, are extrapolations of the parametric form. For example, the fitted temperature of L1544 at its 2.9$\arcmin$ beam, 13.2\,K, represents an average over a region about 6 times larger than the innermost $\sim$5000\,AU, where \citet{Crapsi2007} resolve gas at only a few K, and is therefore not directly comparable to their value. 
We obtain HI abundance profiles with lower central abundances in both cores, similar to the case of B227 reported by \citet{Zuo2018}. Previous HINSA studies \citep{Goldsmith2005, Zuo2018} failed to detect this phenomenon in these two cores, demonstrating the necessity of our \texttt{hinspheres} framework.
The peak HI shells lie at the outermost radius, qualitatively consistent with the inside-out HI-to-H$_2$ transition picture of \citet{Goldsmith2007}.  
For these two cores, the fitted abundance profiles span about three orders of magnitude. Among them, L1544 shows a lower central HI abundance, reaching $0.003\%$. Neither of these two cores shows significant infall.
In the fitting, we simply set $v_{\rm lsr}$ to 0.0 and 7.0 km/s. Combined with the fitted $\delta_v$, we derive HINSA line‐center velocities of $-0.27$ and $6.90$ km/s for the two cores, respectively.
The residuals of the two fits are not dominated by noise. For both cores, uncertainties in the polynomial background extraction introduce additional residuals. Moreover, as shown in Figure~\ref{fig:center_spec_maps}, L1574 exhibits a more elliptical morphology, which leads to a larger residual.

We use the fitted radial density and HI abundance profiles to calculate the
chemical age of each core.  Following \citet{Goldsmith2005}, the age of a shell with
measured HI abundance $f_\mathrm{HI}$ can be obtained by inverting
Equation~\ref{eq:xhi_profile}.  The age is finite only for
$f_\mathrm{HI} > f_\mathrm{eq}$, where
$f_\mathrm{eq} = \zeta_\mathrm{H_2}/(2k'n+\zeta_\mathrm{H_2})$ is the
equilibrium abundance. Shells with
$f_\mathrm{HI} \lesssim f_\mathrm{eq}$ have already reached chemical
equilibrium and yield only a lower limit on the age.

For L1574, all five shells have abundances above the equilibrium
values, yielding chemical ages of $7.68\times10^5$, $1.37\times10^6$, $3.61\times10^6$, $1.01\times10^7$, $3.43\times10^6$~yr, respectively.
Given that each shell of a given core is expected to share a common chemical age, and that a direct arithmetic average of the shell-by-shell ages would be overly weighted by extreme values, we choose not to adopt the simple average ($3.8\times10^6$~yr for L1574) as the derived age. %Furthermore, the shell-by-shell age inversion is highly nonlinear: near the equilibrium abundance, small uncertainties in $f_\mathrm{HI}$ translate into order-of-magnitude uncertainties in the derived age. 
Instead, we fit a single common age $t$ to the entire radial HI abundance profile by minimizing $\chi^2(t) = \sum_i [\log_{10} f_{\mathrm{HI},i} - \log_{10} x_{\mathrm{HI}}(t; n_{{\rm H},i})]^2$ over all shells, where $x_{\mathrm{HI}}(t; n_{{\rm H},i})$ is Equation~\ref{eq:xhi_profile} evaluated at the density of shell $i$.  This logarithmic objective weights every shell equally in relative terms, and $t$ is restricted to the range $10^3$--$10^9$~yr.  The quoted 1$\sigma$ interval is the 16th--84th percentile range of 2000 bootstrap realizations, each of which perturbs the shell abundances by the fit residual rms in log space.  The fit yields a chemical age of $(6.52\pm1.29)\times10^6$~yr (1$\sigma$) for L1574. This value is in good agreement with the $10^{6.5}$–$10^7$~yr range derived by \citet{Goldsmith2005}. For L1544, not all shells are above the equilibrium abundance: the two innermost shells give ages of $1.60\times10^5$ and $1.08\times10^6$~yr, shell~3 has reached chemical equilibrium and yields only a lower limit of $1.67\times10^6$~yr, shell~4 lies far below the equilibrium abundance and gives no useful age constraint, and the outermost shell gives $8.41\times10^6$~yr. Applying the same single‑age fitting method to the L1544 profile gives an overall chemical age of $(8.66\pm1.70)\times10^6$~yr, indicating that the core is still in the HI-to-H$_2$ transition phase.
\section{Discussion}
\label{sec:discussion}

\subsection{Non-Gaussianity of HINSA Profiles}
\label{sec:nongaussianity}

The HINSA absorption profile $\Delta T(v)$ is not intrinsically
Gaussian.  In the multi-layer radiative transfer framework of
\texttt{hinspheres}, each shell absorbs and re-emits at its own spin
temperature, so the effective excitation temperature seen by the
observer is in general a function of velocity.  When the cloud has a
radial temperature gradient and the line-of-sight optical depth is
non-negligible, the resulting $\Delta T(v)$ deviates from a single
Gaussian.  Two independent physical effects contribute to this
non-Gaussianity. 
First, an inward velocity field projects onto the line of sight as a
radial velocity gradient, causing the HINSA to be distributed over a wider velocity range. For the moderate infall fraction adopted here ($f_\mathrm{ff} = 0.1$%,
%consistent with the quiescent cores studied by \citealt{Goldsmith2005}
),
this effect is small. Second, when the optical depth at the core center is large, the warm foreground layers no longer act as absorbers in the radiative transfer. Instead, their own emission fills in the line center, turning the single-dip absorption into a double dip \citep{Goldsmith2007}.

To quantify the degree of non-Gaussianity across evolutionary stages, we
generate noise-free synthetic cubes at ages
between $10^5$ and $5\times10^8$~yr, using the same fiducial cloud
parameters as in Table~\ref{tab:synth_params}.
For each age we extract the central-beam spectrum and
fit a single Gaussian to $\Delta T(v)$ in the velocity range
$|v| \le 3$~km/s, measuring the residual as the normalized RMS
difference $\sqrt{\langle(\Delta T - \Delta T_\mathrm{gauss})^2\rangle}
/ \max(\Delta T)$. We use this residual to represent the deviation of the HINSA profile from a Gaussian.

\begin{figure}[htbp]
  \centering
  \includegraphics[width=\columnwidth]{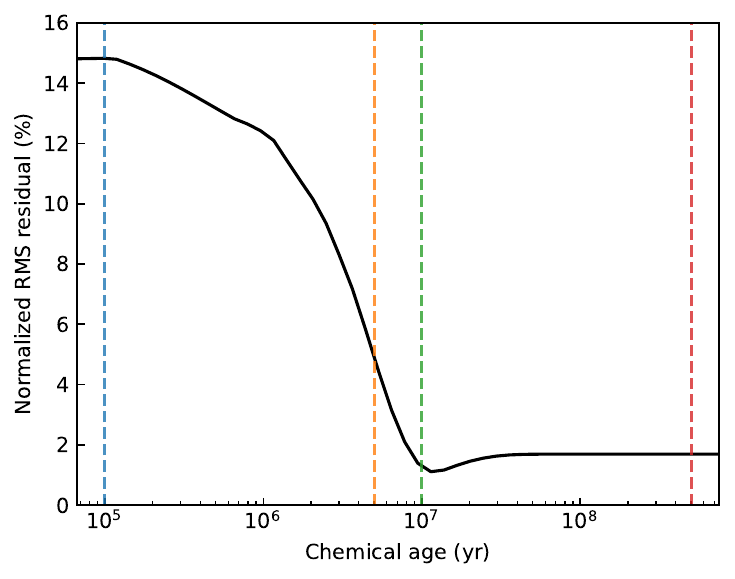}
  \caption{Deviation of the HINSA absorption profile from a single
  Gaussian as a function of chemical age. The normalized RMS residual
  between the central-beam \(\Delta T(v)\) spectrum and its best-fitting
  single Gaussian is used as a measure of this deviation, computed for
  ages from \(10^5\) to \(5\times10^8\)~yr. Vertical dashed lines mark
  the ages of our synthetic cores.}
  \label{fig:gaussian_deviation}
\end{figure}

Figure~\ref{fig:gaussian_deviation} shows the results. For the youngest cores, the saturation-flattened spectra with double-peaked features produce a deviation of $\sim 14.8$\%. As the cloud ages and the HI abundance decreases, the line saturation weakens and the profile becomes more Gaussian. The slight rise beyond $\sim 10^7$\,yr is a normalization artifact: the absolute residual remains nearly constant while the peak $\Delta T$ continues to decrease toward its steady-state value, inflating the ratio. Overall, at ages greater than $\sim 1\times10^7$~yr, the deviation drops to $\sim 1.6$\%, making it possible to describe HINSA with a single Gaussian.

These results show that the younger cores undergoing the HI-to-H$_2$ transition are precisely those whose profiles deviate most strongly from a single Gaussian. Given that these young cores are the key targets for studying the HI-to-H$_2$ transition in molecular clouds, the need for our \texttt{hinspheres} tool is clear. By contrast, for clouds near or at chemical equilibrium, where the HI abundance is low and the profile approaches a Gaussian, the conventional approach remains applicable.

%--------------------------------------------------------------

\subsection{Pixel-by-Pixel Gaussian Fitting}
\label{sec:pixel_gaussian}

Using the synthetic cores in Section~\ref{sec:synthetic}, we compare the \texttt{hinspheres} results with the pixel-by-pixel Gaussian fitting
approach widely used in HINSA
studies~\citep{Li2003}.  For each pixel, the absorption spectrum
$\Delta T(v) = T_\mathrm{bg}(v) - T_\mathrm{obs}(v)$ is fitted with a
Gaussian profile to extract the peak amplitude $A$, central velocity
$v_0$, and velocity dispersion $\sigma_v$.  Unlike
\citet{Li2003}, who have to reconstruct $T_\mathrm{bg}$ by fitting a
polynomial across the narrow absorption feature, the
background is known exactly from the setup for our synthetic cores. This constitutes
a best-case baseline for the pixel-by-pixel Gaussian method. Later
improvements~\citep{Krco2008} use molecular-line templates and
second-derivative spectra to constrain the background, reducing the
ambiguity inherent in the polynomial fit, but the underlying inversion
of the absorption profile, in which the HINSA feature is treated as a
single Gaussian component, remains the same.

Since there is no foreground component for our synthetic cores, we apply two-component radiative transfer fitting to every pixel of these cores. 
Under the two-component radiative transfer comprising only the
background HI emission and the dark cloud, the observed brightness
temperature at velocity $v$ is
\begin{equation}
  T_\mathrm{obs}(v) = T_\mathrm{bg}(v)\,e^{-\tau(v)} +
  T\bigl[1 - e^{-\tau(v)}\bigr],
\end{equation}
where $T$ is the HI spin temperature.  Rearranging gives
the optical depth at the line center,
\begin{equation}
  \tau_0 = -\ln\!\left(
    1 - \frac{A}{T_\mathrm{bg}(v_0) - T}
  \right).
  \label{eq:tau0_inv}
\end{equation}
We adopt $T = 15$~K, a representative value for cold
HINSA gas within the 10--25~K range reported by~\citet{Li2003}.
The HI column density is then obtained from
\begin{equation}
  N_\mathrm{HI} = 1.95\times10^{18}\;\tau_0\;
  \frac{\Delta V}{\mathrm{km/s}}\;
  \frac{T}{\mathrm{K}}
  ~~\mathrm{cm^{-2}},
\end{equation}
with $\Delta V = \sqrt{8\ln 2}\,\sigma_v$ the Gaussian FWHM (\citealt{Li2003}, Equation~13).
This formula is equivalent to the standard
relation $N_\mathrm{HI} = 1.823\times10^{18}\,T
\int\!\tau(v)\,dv$ evaluated for a Gaussian line profile: since
$\int\tau(v)\,dv = \tau_0\,\sigma_v\sqrt{2\pi}$ and
$\Delta V = 2\sqrt{2\ln 2}\,\sigma_v$, the coefficient becomes
$1.823\times10^{18}\times\sqrt{2\pi}/(2\sqrt{2\ln 2}) \sim
1.95\times10^{18}$.

\begin{figure*}[htbp]
  \centering
  \includegraphics[width=0.6\textwidth]{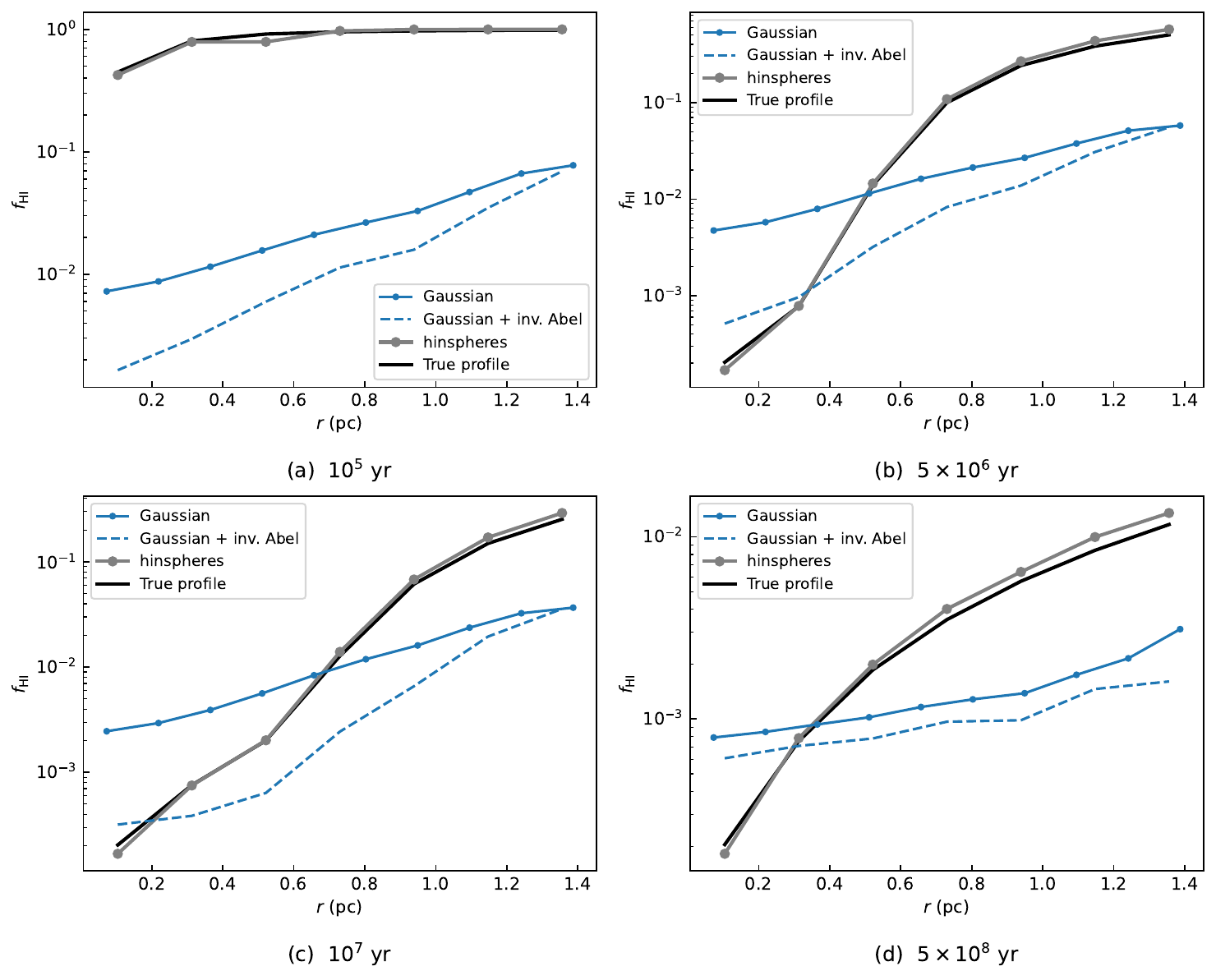}
  \caption{True and fitted radial abundance profiles for the four synthetic cores. Here the 3D volume abundance is \(f_\mathrm{HI}(r)=n_\mathrm{HI}(r)/n_\mathrm{H}(r)\) and the 2D projected column abundance is \(f_\mathrm{HI,2D}(b)=N_\mathrm{HI}(b)/N_\mathrm{H}(b)\), where \(n_\mathrm{H}=n_\mathrm{HI}+2n_\mathrm{H_2}\) and \(N_\mathrm{H}=N_\mathrm{HI}+2N_\mathrm{H_2}\) are the total hydrogen-nucleus volume and column densities.} Black solid lines show the true 3D volume abundance. Gray lines show the 3D volume abundance obtained with \texttt{hinspheres}. Blue solid lines show \(f_\mathrm{HI,2D}(b)\) from pixel-by-pixel single-Gaussian fitting, and blue dashed lines show its inverse-Abel deprojection back to the 3D volume abundance.
  \label{fig:abundance_comparison}
\end{figure*}

Combining the resulting $N_\mathrm{HI}$ map with our derived
$N_\mathrm{H}$ map (see Section~\ref{sec:synthetic}) yields the two-dimensional projected abundance
$f_\mathrm{HI,2D} = N_\mathrm{HI} / N_\mathrm{H}$,
where $N_\mathrm{H} = N_\mathrm{HI} + 2N_{\mathrm{H_2}}$ is the total
hydrogen column. For studies using the pixel-by-pixel 2D fitting approach, it is usually difficult to recover the three-dimensional abundance $f_\mathrm{HI} = n_\mathrm{HI}/n_\mathrm{H}$. Therefore, they typically treat $f_\mathrm{HI,2D}$ as the derived HI abundance for subsequent analysis \citep{Li2003,Krco2008,Krco2010,Tang2020,Zuo2018}.
Using the resulting $f_\mathrm{HI,2D}$ distribution, we derive the radial abundance profile and present it in Figure~\ref{fig:abundance_comparison}. The figure also shows the true 3D radial abundance and the \texttt{hinspheres}-recovered 3D profile for all four evolutionary stages.

As shown in the figure, the pixel-by-pixel Gaussian approach systematically underestimates the peak
abundance and smooths the radial gradient.  For the youngest core, the true
abundance rises from $f_\mathrm{HI} \sim 0.45$ at the innermost
shell to near unity at 1.46~pc. The radially averaged 2D
abundance peaks at only $\sim 0.08$ at the outermost bin, which is a factor
of 13 below the true value at the same radius.  At the oldest
stage, the HI abundance spans two orders of magnitude, from $\sim 10^{-4}$ at the core to $\sim 10^{-2}$ in the outermost shell. The 2D profile, however, is only $\sim 0.001$--$0.003$ and is much flatter than the true profile, overestimating the abundance in the central part and underestimating it in the outermost shell. In real observations, owing to the complex morphology of the core, relatively high noise, and background-fitting errors, this tiny $0.001$--$0.003$ abundance gradient can easily be washed out, making it impossible to observe the ring-like distribution of HI abundance.
By contrast, the fitting results of \texttt{hinspheres} recover the
full radial trend.  Even the innermost
shell of the $10^5$\,yr core, whose line-center opacity is the largest
of any shell, is recovered to within $\sim 5\%$.

The systematic bias of the 2D method has three physical origins.
The first is saturation.  In young cores the inner shells have high HI
density ($n_\mathrm{HI}\sim 5\times10^3$~cm$^{-3}$ for the
$10^5$~yr core), giving a cumulative line-center optical depth
$\tau_0\sim 10^3$ toward the center.  Each layer along the line of
sight both absorbs and re-emits at its own spin temperature.  The cold, optically thick interior absorbs a significant part of the background radiation and lowers the brightness temperature transmitted to the foreground layers. The warm, optically thinner foreground layers can then contribute net emission if their spin temperature exceeds this transmitted brightness temperature, partially filling in the absorption dip. The observed line-center brightness is therefore a mixture of the transmitted background and the foreground emission, rather than the pure absorption signal of the gas. The
pixel-by-pixel method models the cloud as a single isothermal slab
($T=15$~K) and inverts the measured amplitude through
Equation~\ref{eq:tau0_inv}.  Because the warm foreground raises the
observed brightness above what the cold interior alone would produce,
the absorption dip is shallower than the single-slab model expects.
As a result, the maximum $\tau_0$ recoverable from
Equation~\ref{eq:tau0_inv} is $\tau_0^\mathrm{max} \sim 0.96$ at
$T=15$~K ($\sim 0.90$ at 25~K, $\sim 1.1$ at 10~K),
regardless of the true value.
The derived column density therefore saturates at $\sim 4$--$8\times
10^{19}$~cm$^{-2}$, far below the Planck $N_\mathrm{H} \gtrsim
4.5\times10^{20}$~cm$^{-2}$, keeping $f_\mathrm{HI,2D}$ below $\sim 0.1$
at all radii.

Even when the optical depth is moderate and the line shape is close to
Gaussian, the 2D projected abundance is a column-weighted average over
all shells along each line of sight.  For a sight line through the cloud
center, the HI-rich outer layers contribute long path lengths at every
impact parameter, pulling the recovered abundance at inner radii up
toward the high outer-shell values. At
the largest impact parameters the sight line reaches only the outermost
shell and projection contamination is negligible.  The residual
underestimate there is set by the single-$T$
assumption discussed below. 

The pixel-by-pixel method assumes a single representative
$T$ (here 15\,K) for all velocity channels.  In a cloud
with a radial temperature gradient, the spin temperature
of the absorbing column is in fact velocity-dependent: line-center
channels probe deeper, colder layers while the wings sample warmer
outer gas.  The assumed $T$ therefore does not represent
the true radially varying excitation, introducing a systematic offset in
the derived $\tau_0$ and $N_\mathrm{HI}$.  This effect is secondary
compared with saturation and projection mixing, but it contributes to
the residual underestimate at large impact parameters where the optical
depth is moderate. 

The 2D projected abundance can in principle be improved by
deprojection.  The inverse Abel transform converts the projected column
profile of a spherically symmetric cloud into a volume
profile without assuming a particular radial functional form~\citep{Roy2014,Zhang2023}.  We apply it separately to
$N_\mathrm{HI}(b)$ and $N_\mathrm{H}(b)$ and combine into
$f_\mathrm{HI,Abel}(r) = n_\mathrm{HI}(r)/n_\mathrm{H}(r)$
(blue dashed line in Figure~\ref{fig:abundance_comparison}).  The
deprojection does sharpen the radial gradient compared with the flat
2D profile, but it cannot fully recover the true abundance because the
input $N_\mathrm{HI}$ map itself is already compromised by saturation
and the single-$T$ assumption.  For the $10^5$~yr core,
the deprojected inner abundance remains a factor of $\sim 270$ below the
true value.  For the $5\times10^8$~yr core, the recovered profile
stays flat at $\sim 0.0006$--$0.0016$ (shell-averaged relative error
$84\%$), barely changed from the 2D result.  The Abel formula is also
ill-conditioned: the volume density $n(r)$ depends on the derivative $dN/db$ over all
$b > r$, so noise and truncation errors at large radii propagate toward
the center where the signal is weakest.

\begin{figure}[htbp]
  \centering
  \includegraphics[width=\columnwidth]{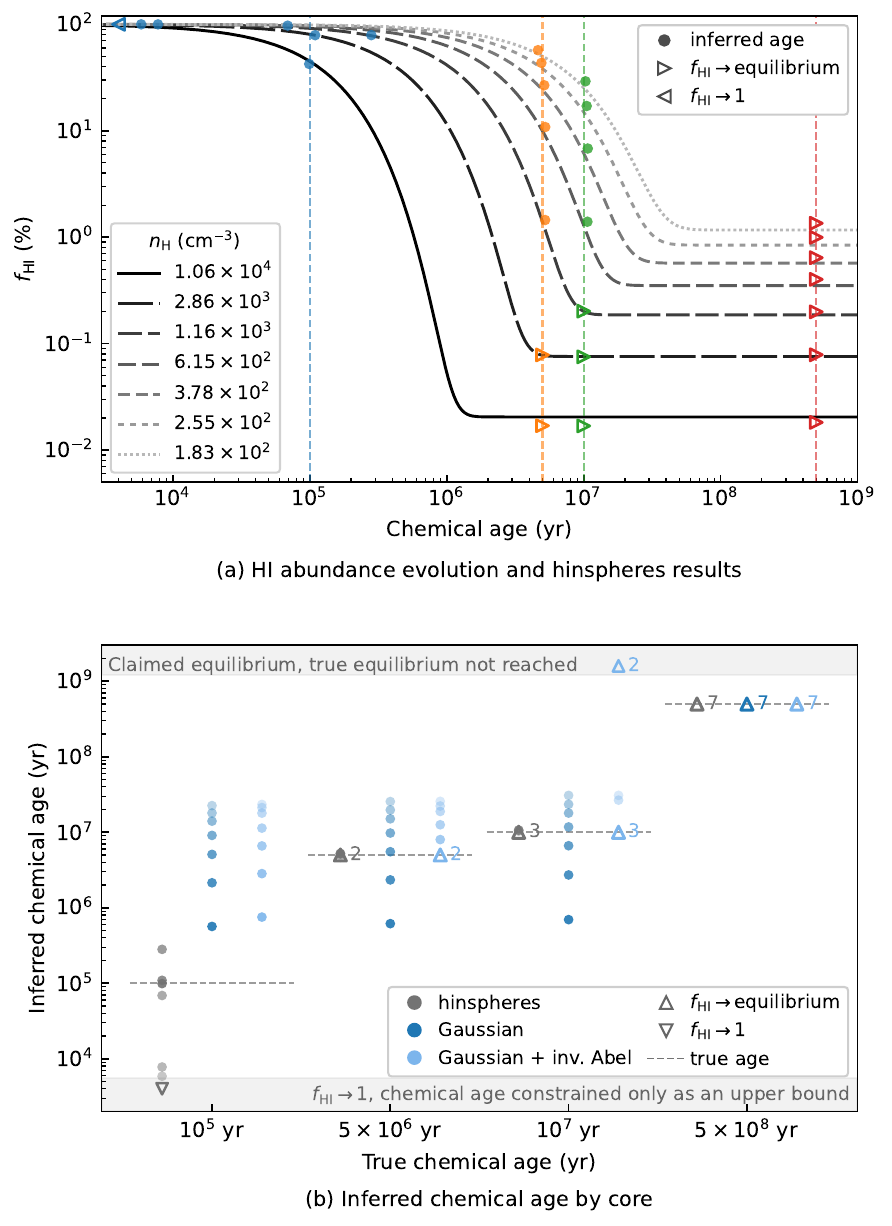}
  \caption{HI abundance evolution and inferred chemical ages derived from \texttt{hinspheres} and two Gaussian-based methods. (a) HI abundance evolution at the densities of the seven shells of our fiducial model. Colored dashed vertical lines mark the true ages of the four synthetic cores, and colored symbols show the \texttt{hinspheres} results. Curves and symbols sharing the same shell index refer to the same shell at different densities: curves use the true density, while fitted abundances reference the \(N_{\rm H}\)-map density adopted in the fit. This leads to a small vertical offset between the curves and the symbols due to the density normalization. Circles denote shells with an invertible age, right-pointing triangles denote shells that have reached equilibrium ($>99.9\%$ of convertible HI converted), and left-pointing triangles denote shells with \(f_{\rm HI} \to 1\) ($<0.1\%$ converted). For clarity, right-pointing triangles for shells that truly reached equilibrium are plotted at their corresponding true ages. (b) Inferred chemical age for each core and method (\texttt{hinspheres}, Gaussian, and Gaussian after inverse Abel correction). All shells of a given core and method share the same horizontal position, with more transparent circles indicating lower-density shells. Shells correctly identified as equilibrium are shown as up-triangles at the core's true age. Shells claimed to be at equilibrium but not truly reached appear in the upper row. Shells with \(f_{\rm HI} \to 1\) appear as down-triangles in the lower row. Dashed horizontal segments mark the true age of each core, and numbers indicate the shell counts when triangle markers overlap.}
  \label{fig:spectra_evolution}
\end{figure}

To assess how these systematic biases translate into errors on the inferred chemical age, we compare the ages derived from \texttt{hinspheres} and the Gaussian-based methods with the true ages of the synthetic cores in Figure~\ref{fig:spectra_evolution}. We consider a shell to have reached equilibrium when at least \(99.9\%\) of its convertible HI at the corresponding density has been converted.
For the \texttt{hinspheres} results in the three older cores, the method reasonably well recovers the corresponding HI abundances regardless of whether a specific shell has reached equilibrium. Consequently, the inferred ages of the various shells within the same core remain similar to the true age. In the outer shells of the oldest core, the fitted abundances lie above the plotted curves (up to \(\sim18\%\) above the true abundances, see Table~\ref{tab:synrec}), yet these shells are still at equilibrium. The reason is that our \texttt{hinspheres} fitting adopts densities derived from our \(N_{\rm H}\)-map. Density errors from fitting this map result in adopted densities being lower than the true values at the outer shells. Since the equilibrium HI abundance scales as \(f_{\rm HI, eq} \sim n_{\rm H}^{-1}\) (see Equation~\ref{eq:xhi_profile}), both the fitted abundance and the equilibrium value are raised. Thus, this offset merely shifts the data point vertically relative to the true-density curve without changing its equilibrium status.
For the \(10^5\)~yr core, because its \(f_{\rm HI}\) is relatively close to 1, \texttt{hinspheres} yields underestimated ages for the three outermost shells, while the ages calculated for the four inner shells remain close to \(10^5\)~yr.
The pixel-by-pixel Gaussian method yields large age errors for the three younger cores, with inferred ages spreading over \(10^6\)--\(3\times10^7\)~yr regardless of their true ages, making the age inference for these cores highly uncertain. The inverse Abel correction narrows this spread slightly. However, for the \(5\times10^8\)~yr core, both Gaussian-based methods correctly identify the fact that the core has reached equilibrium.

In summary, the pixel-by-pixel Gaussian method's bias is dominated by
optical-depth saturation at young stages and by projection mixing at
later stages, with the single-$T$
assumption playing a secondary role throughout.  More fundamentally,
this approach presupposes that the HINSA feature is a pure absorption
dip, whereas in reality each layer both absorbs and re-emits. The conventional approach is
therefore reliable only when the HINSA profile is close to Gaussian,
that is, when the cloud is near chemical equilibrium and the
line-center optical depth is small.  In practice, observers can assess
this by inspecting the absorption profile: a clean, symmetric single-dip
shape indicates that the single-Gaussian inversion is applicable, while
a double-peaked, flattened, or otherwise non-Gaussian profile signals
serious bias.  For such sources, forward multi-shell radiative-transfer
fitting as implemented in \texttt{hinspheres} is necessary.

%--------------------------------------------------------------

\subsection{Applicability and Limitations}
\label{sec:applicability}

The comparisons above demonstrate that multi-shell radiative-transfer
fitting substantially improves over the single-Gaussian inversion,
particularly for young cores where the HINSA profile departs strongly
from a Gaussian shape.  \texttt{hinspheres} achieves this by modeling
the cloud as a spherically symmetric, multi-layer system with parametric
density and temperature profiles.  These assumptions enable efficient,
physically motivated fitting, but they also define the regime in which
the results can be trusted. Real cores may deviate from spherical
geometry, and the density--abundance degeneracy and the sensitivity to
the outer radius impose further constraints.

Our forward model assumes spherical symmetry
(Section~\ref{sec:model}).  L1574 and L1544 approximately satisfy this criterion, though L1574 shows mild ellipticity. Both these cores
are relatively isolated, low-mass cores with
centrally concentrated HINSA structures.  The pipeline recovers
physically meaningful radial profiles for them, and the chemical ages
derived from the abundance profiles are reasonable.  The clearest counter-example is B227,
where the dust-emission peak and the ring-like HINSA center are offset
by several arcmin \citep[Fig.~1]{Zuo2018}, making a single spherically
symmetric model inappropriate. This prevents us from fitting B227 with \texttt{hinspheres}, even though \citet{Zuo2018} has already identified its ring-like HI abundance structure using a pixel-by-pixel single-Gaussian approach. In addition to cases where the dust emission peak and the HI core are offset, cores with pronounced morphological asymmetry (e.g., elongated or filamentary morphology) or those in complex environments (e.g., a one-sided illuminating radiation field) also break the spherical-symmetry assumption and therefore cannot be well described by the parametric model implemented here.

Beyond the geometric requirement, the density--abundance degeneracy is the dominant source of parameter
uncertainty. In Section~\ref{sec:density_T0_grid}, without any density prior,
$\langle\delta f_\mathrm{HI}\rangle$ reaches $182\%$, while even a
loose $\pm1\sigma$ Planck constraint reduces this to $\sim 22\%$.
This degeneracy is intrinsic to the radiative transfer of the HI line.  The density profile must
therefore be anchored by dust continuum, dust extinction, or CO isotopologue data before the
HI abundance can be recovered meaningfully.

Finally, the outer radius must be correctly estimated.
Section~\ref{sec:rout_misest} showed that underestimating $R_\mathrm{out}$
by more than $\sim 18\%$ is catastrophic,
while overestimating by up to $71\%$ produces only moderate biases.  When $R_\mathrm{out}$ is
uncertain, a generous upper bound should therefore be adopted.

If one or more of these conditions is violated, the fit residuals
increase and the derived abundance profiles become unreliable.  For such
sources, axially symmetric or fully three-dimensional models, potentially
coupled with non-LTE line transfer, would be more appropriate.
When all of these conditions are satisfied, however, \texttt{hinspheres} can yield reasonably reliable results.

%%======================================================================
%% 7. Summary
%%======================================================================
\section{Summary}
\label{sec:summary}

We have presented \texttt{hinspheres}, an open-source Python pipeline for
spherically symmetric radiative transfer modeling and fitting of
HINSA.  The package models a cloud as $N$
concentric shells with parametric Plummer density and temperature
profiles, a discrete radial HI abundance distribution, and optional
infall and rigid-body rotation velocity fields.  Layer-by-layer
radiative transfer is solved through a three-dimensional data cube, and
a derivative-free CMA-ES optimizer jointly determines the cloud
structure, kinematics, and HI abundance profile by minimizing a
radially weighted residual between the model and observed data cube.

The pipeline is validated on synthetic HINSA spectra generated from
time-dependent chemical models of four evolutionary stages spanning
$10^5$ to $5\times10^8$~yr.  Application to the real dark cores L1574
and L1544 demonstrates recovery of radially varying HI abundance
profiles and chemical ages consistent with independent estimates.
Key results include:

\begin{enumerate}
  \item With an independent column-density constraint from $N_{\rm H}$
    column density mapping, the Plummer density parameters ($n_0$,
    $r_0$, $\alpha$) are recovered within $\sim10\%$, and the
    CMA-ES fit recovers the remaining 15 parameters (temperature
    structure, seven-shell HI abundance, kinematics) with a mean
    relative error of $\langle\delta_{\mathrm{all}}\rangle =
    4.6$--$6.8\%$ across the four evolutionary stages.
  \item A combined resolution--noise grid with $7\times7$ beam--noise
    combinations and 980 total fits shows that for a core within
    1~kpc observed at FAST resolution, the median recovery error of all
    15 parameters remains below 16\% for RMS $ < 0.2$ K under our
    synthetic conditions, and only exceeds
    50\% for beams of $20\arcmin$ or larger at RMS $\ge0.6$~K, or at
    $30\arcmin$ nearly independently of the noise level.
  \item A systematic study of density and temperature constraint
    schemes demonstrates that the density--abundance degeneracy is the
    dominant source of uncertainty. Without any density prior,
    $\langle\delta f_\mathrm{HI}\rangle$ reaches 182\%, while even a
    loose $\pm1\sigma$ Planck constraint reduces it to $\sim22\%$, and
    fixing the density reduces it further to $\sim9\%$.  A temperature
    prior provides secondary improvement, most valuable when the
    density constraint is only approximate.
  \item Underestimating the total cloud radius $R_\mathrm{out}$ by more
    than $\sim18\%$ is catastrophic,
    while overestimating by up to 71\% produces only moderate biases.
    When $R_\mathrm{out}$ is uncertain, a generous upper bound should
    therefore be adopted.
  \item HINSA profiles of young, HI-rich cores ($\lesssim10^7$~yr)
    deviate strongly from a single Gaussian due to line-center saturation and the
    projected velocity gradient, rendering single-Gaussian analysis
    unreliable for these objects.
  \item A comparison with the conventional pixel-by-pixel single-Gaussian method reveals that the two-dimensional approach suffers from three systematic biases. Optical-depth saturation caps the recoverable column density when the line center is optically thick. Projection mixing causes shells of different abundances to be superposed along each line of sight. The single spin-temperature assumption ignores the velocity-dependent excitation structure. 
    The method remains applicable when the cloud
    is near chemical equilibrium and the HINSA profile is close to
    Gaussian, but for young, HI-rich cores whose profiles deviate
    strongly from a Gaussian, forward multi-shell radiative transfer
    fitting as implemented in \texttt{hinspheres} is necessary.
  \item Application of \texttt{hinspheres} to L1574 and L1544 yields HI abundance profiles
    spanning about three orders of magnitude and single-age chemical
    fits of $(6.52\pm1.29)\times10^6$~yr and
    $(8.66\pm1.70)\times10^6$~yr, respectively, both consistent with
    the inside-out HI-to-H$_2$ transition picture.
\end{enumerate}

\texttt{hinspheres} is publicly available at
\url{https://github.com/olozhika/hinspheres} under the MIT license.

%%======================================================================
%% Acknowledgements
%%======================================================================
\begin{acknowledgments}
  This work is supported by National Natural Science Foundation of China (NSFC) grants Nos. 12588202. We thank Professor Diego Mardones for his insightful discussions.
\end{acknowledgments}

%%======================================================================
%% References
%%======================================================================
\bibliography{references}{}
\bibliographystyle{aasjournalv7}

\end{document}

%% file: synrec_table.tex
% Automatically generated by generate_recovery_table.py -- do not edit manually
\begin{table*}[htbp]
\centering
\caption{Best-fit parameter recovery with $N_{\rm H}$ density constraint for the synthetic cores, L1574, and L1544.}
\label{tab:synrec}
\footnotesize
\setlength{\tabcolsep}{2pt}
\begin{tabular}{lcccccccccccccccccc}
\toprule
  Stage & \multicolumn{7}{c}{$f_{\rm HI}$ (\%)} & \multicolumn{3}{c}{Temperature} & \multicolumn{5}{c}{Kinematics} & \multicolumn{3}{c}{Diagnostics} \\
\cmidrule(lr){2-8} \cmidrule(lr){9-11} \cmidrule(lr){12-16} \cmidrule(lr){17-19}
   & 1 & 2 & 3 & 4 & 5 & 6 & 7 & $T_0$ & $T_1$ & $r_T$ & $\sigma_{\rm turb}$ & $\delta v$ & $f_{\rm ff}$ & $v_{\rm rot}$ & PA & $\langle\delta f_{\rm HI}\rangle$ & $\langle\delta_{\rm all}\rangle$ & Residual \\
   &  &  &  &  &  &  &  & (K) & (K) & (pc) & (km/s) & (km/s) &  & (km/s) & ($^\circ$) & (\%) & (\%) & (K$^2$) \\
\midrule
  \multicolumn{19}{l}{\textbf{$10^5$\,yr}} \\
  True & 44.83 & 80.52 & 91.57 & 95.45 & 97.18 & 98.09 & 98.62 & 10.00 & 58.00 & 0.73 & 0.20 & 0.00 & 0.10 & 0.20 & 45.00 & & & \\
  Fitted & 42.39 & 78.99 & 79.07 & 97.09 & 99.80 & 99.90 & 100.00 & 6.12 & 56.01 & 0.63 & 0.23 & 0.00 & 0.10 & 0.20 & 45.08 & 4.1 & 6.8 & 0.0784 \\
\midrule
  \multicolumn{19}{l}{\textbf{$5\times10^6$\,yr}} \\
  True & 0.02 & 0.08 & 1.39 & 9.96 & 24.13 & 38.28 & 50.10 & 10.00 & 58.00 & 0.73 & 0.20 & 0.00 & 0.10 & 0.20 & 45.00 & & & \\
  Fitted & 0.02 & 0.08 & 1.45 & 10.83 & 26.74 & 43.39 & 57.15 & 9.30 & 57.71 & 0.71 & 0.20 & 0.00 & 0.10 & 0.20 & 45.02 & 7.9 & 4.6 & 0.0285 \\
\midrule
  \multicolumn{19}{l}{\textbf{$10^7$\,yr}} \\
  True & 0.02 & 0.08 & 0.20 & 1.28 & 6.15 & 14.98 & 25.40 & 10.00 & 58.00 & 0.73 & 0.20 & 0.00 & 0.10 & 0.20 & 45.00 & & & \\
  Fitted & 0.02 & 0.08 & 0.20 & 1.40 & 6.81 & 17.07 & 29.14 & 9.21 & 57.94 & 0.72 & 0.20 & 0.00 & 0.11 & 0.20 & 45.10 & 7.6 & 4.9 & 0.0285 \\
\midrule
  \multicolumn{19}{l}{\textbf{$5\times10^8$\,yr}} \\
  True & 0.02 & 0.08 & 0.19 & 0.35 & 0.57 & 0.84 & 1.17 & 10.00 & 58.00 & 0.73 & 0.20 & 0.00 & 0.10 & 0.20 & 45.00 & & & \\
  Fitted & 0.02 & 0.08 & 0.20 & 0.40 & 0.64 & 1.00 & 1.35 & 9.74 & 57.61 & 0.70 & 0.19 & 0.00 & 0.11 & 0.20 & 45.90 & 10.4 & 6.3 & 0.0285 \\
\midrule
  \multicolumn{19}{l}{\textbf{L1574}} \\
  Fitted & 0.03 & 0.15 & 0.22 & 0.22 & 21.79 & - & - & 5.00$^{\dag}$ & 100.00$^{\dag}$ & 0.98 & 0.54 & -0.27 & 0.00 & -0.20 & 57.98 & & & 7.2551 \\
\midrule
  \multicolumn{19}{l}{\textbf{L1544}} \\
  Fitted & 0.00 & 0.02 & 0.03 & 0.03 & 0.77 & - & - & 11.50 & 52.85 & 0.70$^{\dag}$ & 0.00 & -0.10 & 0.00 & -0.27 & 64.73 & & & 2.0356 \\
\bottomrule
\end{tabular}
\vspace{2pt}
{\footnotesize $\dag$~Value at the boundary of the adopted parameter range ($T_0$: 5--30\,K, $T_1$: 10--100\,K, $r_T$: 0.01\,pc--$R_{\rm out}$); such values are formal parameters of the parametric temperature profile rather than temperature measurements (see Section~\ref{sec:application}).}
{\footnotesize Residual: weighted mean of $(T_{\rm obs}-T_{\rm mod})^2$ over the fit window $|v-v_{\rm LSR}|\le3$~km/s with weights proportional to $1/r$, computed within $r\le R_{\rm out}$ (K$^2$). The term beyond $R_{\rm out}$, where the model equals the adopted background by construction, is excluded.}
\end{table*}